\documentclass[twocolumn,dvipsnames]{aastex63}

\usepackage[utf8]{inputenc}
\usepackage{amsmath}
\usepackage{booktabs}
\usepackage{float}

\newcommand{\ddlr}{$d_{\mathrm{DLR}}$ }
\newcommand{\snnz}{$\text{SNN}_{+Z}$}
\newcommand{\snnnoz}{$\text{SNN}_{NoZ}$}
\newcommand{\diffimg}{\texttt{DiffImg}}

\DeclareUnicodeCharacter{FF0C}{é}

\begin{document}
\title{The Dark Energy Survey Supernova Program: Cosmological Biases from Host Galaxy Mismatch of Type Ia Supernovae}

\author[0000-0003-1899-9791]{H.~Qu}
\affiliation{Department of Physics and Astronomy, University of Pennsylvania, Philadelphia, PA 19104, USA}
\author[0000-0003-2764-7093]{M.~Sako}
\affiliation{Department of Physics and Astronomy, University of Pennsylvania, Philadelphia, PA 19104, USA}
\author{M.~Vincenzi}
\affiliation{Department of Physics, Duke University Durham, NC 27708, USA}
\author{C.~S{\'a}nchez}
\affiliation{Departament de Física, Universitat Autònoma de Barcelona, 08193 Bellaterra, Barcelona, Spain}
\affiliation{Institut de Física d’Altes Energies (IFAE), The Barcelona Institute of Science and Technology, Campus UAB, 08193 Bellaterra, Barcelona, Spain}
\author[0000-0001-5201-8374]{D.~Brout}
\affiliation{Department of Astronomy, Boston University, 725 Commonwealth Ave., Boston, MA 02215, USA}
\author[0000-0003-3221-0419]{R.~Kessler}
\affiliation{Department of Astronomy and Astrophysics, University of Chicago, Chicago, IL 60637, USA}
\affiliation{Kavli Institute for Cosmological Physics, University of Chicago, Chicago, IL 60637, USA}
\author{R.~Chen}
\affiliation{Department of Physics, Duke University Durham, NC 27708, USA}
\author[0000-0002-4213-8783]{T.~Davis}
\affiliation{School of Mathematics and Physics, The University of Queensland, Brisbane QLD 4072 Australia}
\author[0000-0002-1296-6887]{L.~Galbany}
\affiliation{Institut d'Estudis Espacials de Catalunya (IEEC), 08034 Barcelona, Spain}
\affiliation{Institute of Space Sciences (ICE, CSIC),  Campus UAB, Carrer de Can Magrans, s/n,  08193 Barcelona, Spain}
\author{L.~Kelsey}
\affiliation{Institute of Cosmology and Gravitation, University of Portsmouth, Portsmouth, PO1 3FX, UK}
\affiliation{School of Physics and Astronomy, University of Southampton,  Southampton, SO17 1BJ, UK}
\author[0000-0001-6633-9793]{J.~Lee}
\affiliation{Department of Physics and Astronomy, University of Pennsylvania, Philadelphia, PA 19104, USA}
\author[0000-0003-1731-0497]{C.~Lidman}
\affiliation{Centre for Gravitational Astrophysics, College of Science, The Australian National University, ACT 2601, Australia}
\affiliation{The Research School of Astronomy and Astrophysics, Australian National University, ACT 2601, Australia}
\author[0000-0002-8012-6978]{B.~Popovic}
\affiliation{Department of Physics, Duke University Durham, NC 27708, USA}
\author[0000-0002-1873-8973]{B.~Rose}
\affiliation{Department of Physics, Duke University Durham, NC 27708, USA}
\author[0000-0002-4934-5849]{D.~Scolnic}
\affiliation{Department of Physics, Duke University Durham, NC 27708, USA}
\author[0000-0002-3321-1432]{M.~Smith}
\affiliation{Univ Lyon, Univ Claude Bernard Lyon 1, CNRS, IP2I Lyon / IN2P3, IMR 5822, F-69622 Villeurbanne, France}
\author[0000-0001-9053-4820]{M.~Sullivan}
\affiliation{School of Physics and Astronomy, University of Southampton,  Southampton, SO17 1BJ, UK}
\author{P.~Wiseman}
\affiliation{School of Physics and Astronomy, University of Southampton,  Southampton, SO17 1BJ, UK}
\author{T.~M.~C.~Abbott}
\affiliation{Cerro Tololo Inter-American Observatory, NSF's National Optical-Infrared Astronomy Research Laboratory, Casilla 603, La Serena, Chile}
\author{M.~Aguena}
\affiliation{Laborat\'orio Interinstitucional de e-Astronomia - LIneA, Rua Gal. Jos\'e Cristino 77, Rio de Janeiro, RJ - 20921-400, Brazil}
\author{O.~Alves}
\affiliation{Department of Physics, University of Michigan, Ann Arbor, MI 48109, USA}
\author{D.~Bacon}
\affiliation{Institute of Cosmology and Gravitation, University of Portsmouth, Portsmouth, PO1 3FX, UK}
\author{E.~Bertin}
\affiliation{CNRS, UMR 7095, Institut d'Astrophysique de Paris, F-75014, Paris, France}
\affiliation{Sorbonne Universit\'es, UPMC Univ Paris 06, UMR 7095, Institut d'Astrophysique de Paris, F-75014, Paris, France}
\author[0000-0002-8458-5047]{D.~Brooks}
\affiliation{Department of Physics \& Astronomy, University College London, Gower Street, London, WC1E 6BT, UK}
\author{D.~L.~Burke}
\affiliation{Kavli Institute for Particle Astrophysics \& Cosmology, P. O. Box 2450, Stanford University, Stanford, CA 94305, USA}
\affiliation{SLAC National Accelerator Laboratory, Menlo Park, CA 94025, USA}
\author[0000-0003-3044-5150]{A.~Carnero~Rosell}
\affiliation{Instituto de Astrofisica de Canarias, E-38205 La Laguna, Tenerife, Spain}
\affiliation{Laborat\'orio Interinstitucional de e-Astronomia - LIneA, Rua Gal. Jos\'e Cristino 77, Rio de Janeiro, RJ - 20921-400, Brazil}
\affiliation{Universidad de La Laguna, Dpto. Astrofísica, E-38206 La Laguna, Tenerife, Spain}
\author[0000-0002-3130-0204]{J.~Carretero}
\affiliation{Institut de F\'{\i}sica d'Altes Energies (IFAE), The Barcelona Institute of Science and Technology, Campus UAB, 08193 Bellaterra (Barcelona) Spain}
\author{L.~N.~da Costa}
\affiliation{Laborat\'orio Interinstitucional de e-Astronomia - LIneA, Rua Gal. Jos\'e Cristino 77, Rio de Janeiro, RJ - 20921-400, Brazil}
\author{M.~E.~S.~Pereira}
\affiliation{Hamburger Sternwarte, Universit\"{a}t Hamburg, Gojenbergsweg 112, 21029 Hamburg, Germany}
\author[0000-0002-8357-7467]{H.~T.~Diehl}
\affiliation{Fermi National Accelerator Laboratory, P. O. Box 500, Batavia, IL 60510, USA}
\author{P.~Doel}
\affiliation{Department of Physics \& Astronomy, University College London, Gower Street, London, WC1E 6BT, UK}
\author{S.~Everett}
\affiliation{Jet Propulsion Laboratory, California Institute of Technology, 4800 Oak Grove Dr., Pasadena, CA 91109, USA}
\author{I.~Ferrero}
\affiliation{Institute of Theoretical Astrophysics, University of Oslo. P.O. Box 1029 Blindern, NO-0315 Oslo, Norway}
\author[0000-0003-4079-3263]{J.~Frieman}
\affiliation{Fermi National Accelerator Laboratory, P. O. Box 500, Batavia, IL 60510, USA}
\affiliation{Kavli Institute for Cosmological Physics, University of Chicago, Chicago, IL 60637, USA}
\author[0000-0002-9370-8360]{J.~Garc\'ia-Bellido}
\affiliation{Instituto de Fisica Teorica UAM/CSIC, Universidad Autonoma de Madrid, 28049 Madrid, Spain}
\author[0000-0002-3730-1750]{G.~Giannini}
\affiliation{Institut de F\'{\i}sica d'Altes Energies (IFAE), The Barcelona Institute of Science and Technology, Campus UAB, 08193 Bellaterra (Barcelona) Spain}
\author[0000-0003-3270-7644]{D.~Gruen}
\affiliation{University Observatory, Faculty of Physics, Ludwig-Maximilians-Universit\"at, Scheinerstr. 1, 81679 Munich, Germany}
\author{R.~A.~Gruendl}
\affiliation{Center for Astrophysical Surveys, National Center for Supercomputing Applications, 1205 West Clark St., Urbana, IL 61801, USA}
\affiliation{Department of Astronomy, University of Illinois at Urbana-Champaign, 1002 W. Green Street, Urbana, IL 61801, USA}
\author[0000-0003-0825-0517]{G.~Gutierrez}
\affiliation{Fermi National Accelerator Laboratory, P. O. Box 500, Batavia, IL 60510, USA}
\author{S.~R.~Hinton}
\affiliation{School of Mathematics and Physics, University of Queensland,  Brisbane, QLD 4072, Australia}
\author{D.~L.~Hollowood}
\affiliation{Santa Cruz Institute for Particle Physics, Santa Cruz, CA 95064, USA}
\author[0000-0002-6550-2023]{K.~Honscheid}
\affiliation{Center for Cosmology and Astro-Particle Physics, The Ohio State University, Columbus, OH 43210, USA}
\affiliation{Department of Physics, The Ohio State University, Columbus, OH 43210, USA}
\author[0000-0001-5160-4486]{D.~J.~James}
\affiliation{Center for Astrophysics $\vert$ Harvard \& Smithsonian, 60 Garden Street, Cambridge, MA 02138, USA}
\author[0000-0003-0120-0808]{K.~Kuehn}
\affiliation{Australian Astronomical Optics, Macquarie University, North Ryde, NSW 2113, Australia}
\affiliation{Lowell Observatory, 1400 Mars Hill Rd, Flagstaff, AZ 86001, USA}
\author[0000-0002-1134-9035]{O.~Lahav}
\affiliation{Department of Physics \& Astronomy, University College London, Gower Street, London, WC1E 6BT, UK}
\author[0000-0003-0710-9474]{J.~L.~Marshall}
\affiliation{George P. and Cynthia Woods Mitchell Institute for Fundamental Physics and Astronomy, and Department of Physics and Astronomy, Texas A\&M University, College Station, TX 77843,  USA}
\author[0000-0001-9497-7266]{J. Mena-Fern{\'a}ndez}
\affiliation{Centro de Investigaciones Energ\'eticas, Medioambientales y Tecnol\'ogicas (CIEMAT), Madrid, Spain}
\author[0000-0002-1372-2534]{F.~Menanteau}
\affiliation{Center for Astrophysical Surveys, National Center for Supercomputing Applications, 1205 West Clark St., Urbana, IL 61801, USA}
\affiliation{Department of Astronomy, University of Illinois at Urbana-Champaign, 1002 W. Green Street, Urbana, IL 61801, USA}
\author[0000-0002-6610-4836]{R.~Miquel}
\affiliation{Instituci\'o Catalana de Recerca i Estudis Avan\c{c}ats, E-08010 Barcelona, Spain}
\affiliation{Institut de F\'{\i}sica d'Altes Energies (IFAE), The Barcelona Institute of Science and Technology, Campus UAB, 08193 Bellaterra (Barcelona) Spain}
\author[0000-0003-2120-1154]{R.~L.~C.~Ogando}
\affiliation{Observat\'orio Nacional, Rua Gal. Jos\'e Cristino 77, Rio de Janeiro, RJ - 20921-400, Brazil}
\author[0000-0002-6011-0530]{A.~Palmese}
\affiliation{Department of Physics, Carnegie Mellon University, Pittsburgh, Pennsylvania 15312, USA}
\author[0000-0001-9186-6042]{A.~Pieres}
\affiliation{Laborat\'orio Interinstitucional de e-Astronomia - LIneA, Rua Gal. Jos\'e Cristino 77, Rio de Janeiro, RJ - 20921-400, Brazil}
\affiliation{Observat\'orio Nacional, Rua Gal. Jos\'e Cristino 77, Rio de Janeiro, RJ - 20921-400, Brazil}
\author[0000-0002-2598-0514]{A.~A.~Plazas~Malag\'on}
\affiliation{Department of Astrophysical Sciences, Princeton University, Peyton Hall, Princeton, NJ 08544, USA}
\author{M.~Raveri}
\affiliation{Department of Physics, University of Genova and INFN, Via Dodecaneso 33, 16146, Genova, Italy}
\author[0000-0002-9646-8198]{E.~Sanchez}
\affiliation{Centro de Investigaciones Energ\'eticas, Medioambientales y Tecnol\'ogicas (CIEMAT), Madrid, Spain}
\author[0000-0002-1831-1953]{I.~Sevilla-Noarbe}
\affiliation{Centro de Investigaciones Energ\'eticas, Medioambientales y Tecnol\'ogicas (CIEMAT), Madrid, Spain}
\author[0000-0001-6082-8529]{M.~Soares-Santos}
\affiliation{Department of Physics, University of Michigan, Ann Arbor, MI 48109, USA}
\author[0000-0002-7047-9358]{E.~Suchyta}
\affiliation{Computer Science and Mathematics Division, Oak Ridge National Laboratory, Oak Ridge, TN 37831}
\author[0000-0003-1704-0781]{G.~Tarle}
\affiliation{Department of Physics, University of Michigan, Ann Arbor, MI 48109, USA}
\author{N.~Weaverdyck}
\affiliation{Department of Physics, University of Michigan, Ann Arbor, MI 48109, USA}
\affiliation{Lawrence Berkeley National Laboratory, 1 Cyclotron Road, Berkeley, CA 94720, USA}

\correspondingauthor{Helen Qu}
\email{helenqu@sas.upenn.edu}

\collaboration{100}{(DES Collaboration)}
\begin{abstract}
Redshift measurements, primarily obtained from host galaxies, are essential for inferring cosmological parameters from type Ia supernovae (SNe Ia). Matching SNe to host galaxies using images is non-trivial, resulting in a subset of SNe with mismatched hosts and thus incorrect redshifts. We evaluate the host galaxy mismatch rate and resulting biases on cosmological parameters from simulations modeled after the Dark Energy Survey 5-Year (DES-SN5YR) photometric sample. For both DES-SN5YR data and simulations, we employ the directional light radius method for host galaxy matching. In our SN Ia simulations, we find that 1.7\% of SNe are matched to the wrong host galaxy, with redshift difference between the true and matched host of up to 0.6. Using our analysis pipeline, we determine the shift in  the dark energy equation of state parameter ($\Delta w$) due to including SNe with incorrect host galaxy matches.  For SN Ia-only simulations, we find $\Delta w = 0.0013 \pm 0.0026$ with constraints from the cosmic microwave background (CMB). Including core-collapse SNe and peculiar SNe Ia in the simulation, we find that $\Delta w$ ranges from 0.0009 to 0.0032 depending on the photometric classifier used. This bias is an order of magnitude smaller than the expected total uncertainty on $w$ from the DES-SN5YR sample of $\sim 0.03$.  We conclude that the bias on $w$ from host galaxy mismatch is much smaller than the uncertainties expected from the DES-SN5YR sample, but we encourage further studies to reduce this bias through better host-matching algorithms or selection cuts.
\vspace{0.25in}
\end{abstract}

\section{Introduction}
Type Ia supernovae (SNe Ia) enabled the discovery of accelerating cosmic expansion \citep{perlmutter, riess} and have since been an important probe of the dark energy thought to cause it. To constrain cosmology, each SN Ia must have an accurate estimate of cosmological redshift as well as physical distance. Distance estimates are obtained from the standardized luminosities of SNe Ia, earning them the title of \textit{standardizable candles}, but redshift is difficult to determine without spectroscopy. The depth of modern photometric surveys has resulted in orders of magnitude more photometrically observed SNe than can be followed up spectroscopically. Thus, the vast majority of SNe Ia used for estimation of cosmological parameters are assigned the redshift of their matched host galaxies. Moreover, galaxy redshifts are more precise than redshifts measured from SN spectroscopy due to the broadened features and phase-dependent nature of SN spectra.

\begin{figure}
    \includegraphics[scale=0.32,trim={2cm 0cm 0cm 0cm}]{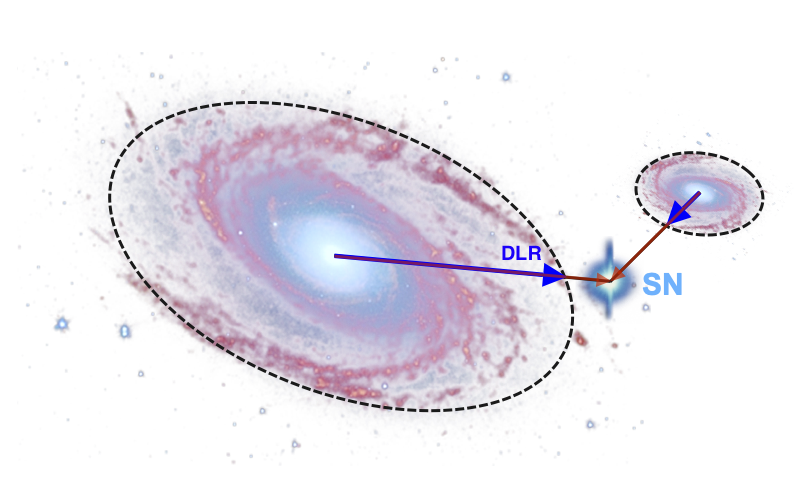}
    \centering
    \caption{An illustration adapted from \cite{gupta} showing an example of a challenging host galaxy matching problem. The supernova (labeled ``SN") is closer in angular separation (red arrows) to the smaller galaxy on the right, but it is closer to the edge of the larger galaxy on the left. The directional light radius (DLR) of each galaxy is shown in the blue arrows. According to the DLR method, the most likely host galaxy is the one with minimal $d_{\mathrm{DLR}}$ value, or ratio of the angular separation to the DLR, which would correctly identify the larger galaxy on the left as the more likely host.}
    \label{fig:host-confusion}
\end{figure}

This work investigates the impact of mismatched host galaxies and the resulting incorrect SN redshifts on the measurement of cosmological parameters. Using images alone, host galaxy matching is a nontrivial problem when there are multiple galaxies near the SN location. Two-dimensional images have little distance information, and galaxies in a crowded field can be difficult to disentangle without distance measurements to each. In addition, the large scale structure of the universe dictates that many galaxies occur in pairs, groups, or clusters, making it difficult to determine which galaxy is the host. Finally, some ``hostless" SNe explode in extremely faint or distant galaxies that fall below the threshold of detection, even in the deep coadded images created for this work described in Section~\ref{subsec:hostgal-catalog}. These SNe could be incorrectly matched to brighter, nearby galaxies that are close in projected distance and thus assigned an incorrect redshift. Figure~\ref{fig:host-confusion} illustrates a challenging example of host galaxy matching, where the larger and more likely host galaxy is further in terms of SN-galaxy separation than the smaller galaxy on the right.

Though automated methods for host galaxy matching have been in use since the SuperNova Legacy Survey (SNLS) analysis \citep[][S06]{sullivan_2006}, accurate characterization of systematic uncertainties such as the effect of mismatched host galaxies has become more pressing with the advent of wider and deeper SN surveys. Future surveys such as the Legacy Survey of Space and Time (LSST) at the Vera Rubin Observatory will observe hundreds of thousands more SNe in the coming decade \citep{lsst_book}, further shrinking statistical uncertainties and necessitating accurate measurements of systematic uncertainties. This work represents the first thorough exploration of systematics related to the host galaxy mismatch problem and its effect on cosmological parameter estimates as part of an ongoing cosmology analysis.


A commonly used method for matching SNe with a host galaxy is the directional light radius (DLR) method, initially developed to characterize host properties for the SNLS survey by S06 and tested extensively on simulations by \citet{gupta}. 
More details about the DLR method can be found in Section~\ref{subsec:dlr}. Recently, \citet[][P20]{popovic2020} performed the first retrospective estimate of cosmological biases resulting from host galaxy mismatches by applying the DLR method on simulations based on SDSS data. P20 found a mismatch rate of 0.6\% with a resulting bias on $w$ of $\Delta w = 0.0007$. 

Recently, a number of promising alternative approaches for host matching have emerged, including DELIGHT \citep{delight} and GHOST \citep{ghost}. DELIGHT introduces a deep learning-based approach to host galaxy identification in which a convolutional neural network is trained on transient images to predict a 2-dimensional vector connecting the transient position with the position of its host galaxy center. GHOST uses a novel gradient ascent method for host galaxy identification that was shown to produce accurate matches in situations where DLR was unable to identify a match. Though this work did not explore these novel approaches, they certainly merit consideration for future SNe Ia cosmological analyses.

In this work, we focus on understanding the cosmological biases arising from host galaxy mismatch using the DLR method on the full Dark Energy Survey 5-Year photometric SNe Ia sample~\citep[DES-SN5YR;][]{vincenzi2021, photo_Ia_des}. Section~\ref{sec:methods} details the DLR host matching algorithm and an overview of our analysis strategy. Section~\ref{sec:data} describes the DES-SN5YR data sample along with the host galaxy catalog used in this analysis. Section~\ref{sec:sims} reviews the simulations used to model the host mismatch rate and characterize its effects on the cosmological parameters, and Section~\ref{subsec:params} describes the methods and results used to ensure consistency between the DES-SN5YR data and our simulations. Section~\ref{sec:cosmo} reviews the framework used for cosmological parameter estimation in this work, and Section~\ref{sec:results} describes the cosmological biases resulting from host galaxy mismatch.

\section{Host Galaxy Matching}
\label{sec:methods}

\subsection{The DLR Method}
\label{subsec:dlr}
The directional light radius (DLR) method for host galaxy matching, developed in S06, involves computing a dimensionless distance ($d_{\mathrm{DLR}}$) for each potential host galaxy measured between the SN position and the centroid of the galaxy normalized by the galaxy size in the direction of the SN. Explicitly, $d_{\mathrm{DLR}}$ is defined as
\begin{equation}
d_{\mathrm{DLR}} = \frac{\mathrm{\Delta \theta}}{\mathrm{DLR}}
\end{equation}
where $\Delta \theta$ is the angular separation (arcsec) between the galaxy centroid and the SN, and DLR (arcsec) is the radius of the galaxy in the direction of the SN. We require all galaxies in our catalog (see Section~\ref{subsec:hostgal-catalog}) to be well modeled by ellipses parameterized by a semi-major axis $a$, a semi-minor axis $b$, and a position angle $\phi$, which is defined relative to the positive RA axis. Thus, the DLR value for each SN-galaxy pair is computed as follows:

\begin{equation}
    \textrm{DLR} = \frac{ab}{\sqrt{(a \,\textrm{sin} \, \phi)^2 + (b \,\textrm{cos} \, \phi)^2}}.
\end{equation}

For our DLR calculations, the values for the $a, b, \phi$ parameters are the \texttt{A\_{IMAGE}}, \texttt{B\_{IMAGE}} (converted into arcsec), and \texttt{THETA\_{IMAGE}} parameters output by Source Extractor \citep[][\texttt{sextractor}]{psfex} using the coadded $r+i+z$ detection image. The galaxy with the lowest $d_{\mathrm{DLR}}$ value is chosen as the host galaxy.

\subsection{Analysis Overview}
\label{}
We quantify the cosmological biases resulting from host galaxy mismatch by generating two sets of simulated SNe, one with and one without host galaxy mismatches, and comparing the fitted cosmological parameters. Details about the DES data can be found in Section~\ref{sec:data} and the simulations are described in Section~\ref{sec:sims}. 

To ensure that our results from simulations are applicable to real DES data, we show consistency between simulations and data across relevant parameter distributions, following \cite{popovic2020}: 
\begin{itemize}
    \item angular SN-galaxy separation ($\Delta \theta$),
    \item directional light radius (DLR),
    \item the DLR-normalized SN-galaxy separation, $d_{\mathrm{DLR}} = \frac{\Delta \theta}{\mathrm{DLR}}$
    \item $r$-band host galaxy magnitude ($m_{r,\mathrm{gal}}$),
    \item the ratio between smallest and second smallest $d_{\mathrm{DLR}}$ values, $r_{\mathrm{DLR}} = \frac{d_{\mathrm{DLR, HOSTGAL1}}}{d_{\mathrm{DLR, HOSTGAL2}}}$, where $d_{\mathrm{DLR, HOSTGAL}\{i\}}$ is the $i^{\text{th}}$ smallest $d_{\mathrm{DLR}}$ value, i.e. the $d_{\mathrm{DLR}}$ value of the $i^{\text{th}}$ most likely host galaxy.
\end{itemize}
 Further description of these parameters and the results of our consistency checks can be found in Section~\ref{subsec:params}. Host matching is performed on one set of simulated SNe to model mismatches as well as the DES-SN5YR data with the DLR method using the catalog of galaxies generated from deep DES imaging stacks, as described in Section~\ref{subsec:hostgal-catalog}. Finally, we estimate biases by comparing the fitted cosmological parameters from simulations with matched hosts and an identical set of simulations with true hosts, i.e. those assigned by the simulation.

\section{Data and host galaxy catalog}
\label{sec:data}

\subsection{Dark Energy Survey 5-Year Photometric Sample}
DES is an optical imaging survey designed to deliver precision cosmological results and constraints on dark energy by combining the probing power of weak gravitational lensing, baryon acoustic oscillations, galaxy clusters, and SNe Ia \citep{des_combined_probes}. DES imaged 5000 $\text{deg}^2$ of the southern sky for 6 years using the Dark Energy Camera \citep{decam} mounted on the 4m Blanco Telescope at the Cerro Tololo Inter-American Observatory. The time-domain survey component of the DES survey strategy covers a smaller area on the sky (10 supernova fields covering 27 $\text{deg}^2$), but exposures were repeated approximately weekly over the course of the survey. Eight of the ten fields were surveyed to a depth of $\sim 23.5$ mag per visit (‘shallow fields’), and the remaining two to a deeper limit of $\sim 24.5$ mag per visit (‘deep fields’), thus extending the SN detection limit to $z \sim 1.2$. Transient identification from images was performed using difference imaging \citep[\diffimg;][hereafter K15]{diffimg}. Spectroscopic follow-up of SN Ia candidates as well as their host galaxies was performed as described in \citet{smith2020,lidman2020}.

\subsection{Host Galaxy Catalog}
\label{subsec:hostgal-catalog}
To ensure sufficient depth and density of potential host galaxies as well as consistency between the galaxy catalogs used for real data matching and simulation, we produce a deep galaxy catalog by coadding DES images, identifying sources, and estimating photometric redshifts and galaxy parameters for each source.

The coadd procedure is similar to those described in \citet{wiseman2020} with several updates.  First, we apply stricter selection requirements (cuts) on the quality of the single-epoch images.  Images with effective exposure time ratio \footnote{$\tau_{\mathrm{eff}} = \Big( \frac{FWHM_{fid}}{FWHM} \Big) ^2 \Big( \frac{B_{fid}}{B} \Big) F_{trans}$, where $FWHM$ and $FWHM_{fid}$ are the measured and fiducial PSF full width half max, respectively; $B$ and $B_{fid}$ are the measured and fiducial sky background; $F_{trans}$ is the atmospheric transmission relative to a nearly clear night. See \citet{morganson} for details.} $\tau_{\mathrm{eff}}$ $\le 0.3$ and those with point spread function full width at half maximum (PSF FWHM) $\ge 1.3\arcsec, 1.2\arcsec, 1.1\arcsec, 1.0\arcsec$ in {\it griz}, respectively, are excluded to ensure higher quality images across the focal plane and to mitigate source confusion.  Second, each image is scaled to a common zeropoint using the same framework adopted in \diffimg\ (see K15).  Third, rather than excluding images from the season in which the supernova candidate was discovered, we instead use all images from the 5-year survey and perform a median coaddition with \texttt{swarp} \citep{psfex}. Although median coaddition is not statistically optimal, it loses only $\sim 0.1$ mag in depth compared to other weighted average methods and is a more robust way to exclude light from transients and image artifacts when adding hundreds of images.  Finally, we determine the PSF model of the coadded images with \texttt{psfex} \citep{psfex} using our tertiary standard stars.  The PSFs are used by \texttt{sextractor} \citep{psfex} to fit for the true (unblurred) Sérsic profiles and derive their parameters, which are necessary for the placement of simulated SNe in their host galaxies.


Sources from all 10 DES SN fields are identified using \texttt{sextractor}. Magnitudes are corrected for Milky Way dust using $E(B-V)$ values from \citet{sfd} and extinction coefficients for DECam filters from \citet{schlafly2011} and assuming $R_V = 3.1$.  These coefficients are $A/E(B-V) = 3.237, 2.176, 1.595,$ and $1.217$ in DECam $griz$, respectively.

The galaxies were matched to a spectroscopic redshift catalog compiled from multiple surveys including OzDES \citep{lidman2020}, SDSS \citep{sdss_followup}, 6dF \citep{6df}, ATLAS \citep{atlas}, GAMA \citep{gama}, VVDS \citep{vvds}, VIPERS \citep{vipers}, ACES \citep{aces}, DEEP2 \citep{deep2}, 2dFGRS \citep{2dfgrs}, UDS/VANDELS \citep{uds}, and PRIMUS \citep{primus}. Among the 7.8 million galaxies, a total of 124,824 have secure spectroscopic redshifts. 
Although we require a spectroscopic redshift for the DES data, the simulation needs a complete galaxy catalog to accurately model mis-matches. We therefore use photometric redshifts when spectroscopic redshifts are not available. 

\begin{figure}
    \includegraphics[scale=0.35,trim={0cm 0cm 0cm 0cm}]{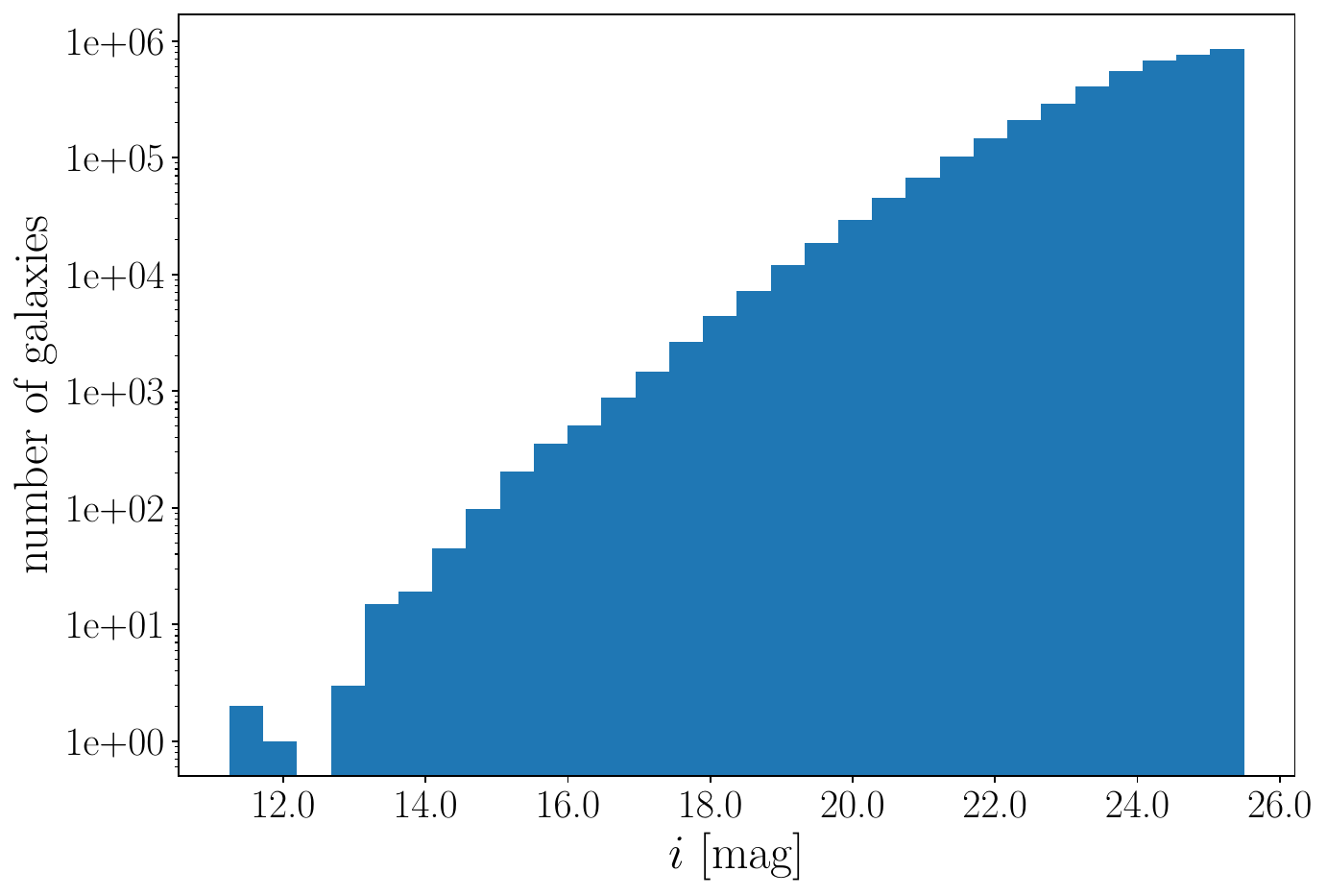}
    \centering
    \caption{$i$-band \texttt{MAG\_MODEL} magnitude distribution of all galaxies in our deep DES galaxy catalog with a spectroscopic or photometric redshift estimate. The catalog cuts off at $i \sim 26$ because photometric redshifts are unavailable for galaxies fainter than $i=25.5$ mag.}
    \label{fig:mag_dist}
\end{figure}

For the vast majority of galaxies without a spectroscopic redshift, we estimate photometric redshifts of galaxies that are brighter than $i=25.5$~mag using the method described in Section~\ref{subsec:photo-z}.  Sources fainter than this limit do not have reliable photo-$z$ estimates and are not included in the galaxy catalog. Figure~\ref{fig:mag_dist}  shows the magnitude distributions of the remaining sources.


\subsubsection{Host Galaxy Photometric Redshifts}
\label{subsec:photo-z}

Host galaxy photometric redshift estimation is performed independently for the deep and shallow SN fields, consisting of two and eight fields, respectively, as described in Section~\ref{subsec:hostgal-catalog}. For each galaxy in each field, we have a set of $griz$ photometric flux measurements and their corresponding uncertainties. Using this data, we train a \emph{Self-Organizing Map} (SOM) to characterize and discretize the photometric space of host galaxies, using the SOM algorithm described in the Appendix of \citet{Sanchez2020}. This SOM algorithm uses unsupervised learning to project the 4-dimensional photometric data ($griz$) onto a lower-dimensional grid, in our case a 2-dimensional grid, while attempting to preserve the topology of the 4-dimensional space. This means that similar objects in the 4-D space will be grouped together in the SOM, enabling a visual understanding of features. 

The particular SOM algorithm used in this work differs from the SOM algorithm used in previous DES analyses \citep[such as][]{y3-sompz, y3-2x2ptaltlenssompz} in order to improve the classification of galaxies with low- and modest-S/N photometry, which is relevant given the faint nature of many SN host galaxies. First, we alter the distance metric used by the SOM algorithm to incorporate flux uncertainties. Next, we include flux information, not just colors, and we do not impose periodic boundary conditions on the map. For the application in this work, we build a SOM of size $22 \times 22$, with a total of 484 cells. These alterations are described in detail in \citet{Sanchez2020}.

After the SOM is constructed, we use the subset of SN host galaxies with spectroscopic redshifts to populate the SOM and compute the redshift distribution of each SOM cell. The spectroscopic subset for deep SN fields has a total of 45,937 galaxies, while the one for shallow SN fields has a total of 78,887 galaxies. Once we populate the SOMs with redshift information, we assign to each galaxy the redshift distribution of the cell it is assigned to. Even if these spectrosopic subsets provide a good coverage of the corresponding photometric spaces, some SOM cells do not contain redshift information and therefore we do not estimate redshifts for galaxies in them. This is the case for 3.9\% of galaxies in the deep SN fields, and 6.7\% of galaxies in the shallow SN fields. 

To validate this procedure, we split the spectroscopic samples into separate training and validation samples, with a random 90\% of galaxies going into training and the remaining 10\% into validation. Next, we compare the estimated and spectroscopic redshifts for galaxies in the validation sample to assess the quality of the reconstruction. For this purpose, and to enable comparisons with previous DES estimates, we use photo-$z$ metrics from \citet{Sanchez2014}.  We construct the $\Delta z \equiv z_{phot} - z_{spec}$ distribution, and we compute two different metrics: 
\begin{enumerate}
    \item We estimate the photo-$z$ precision by calculating the 68-percentile width $\sigma_{68}$ of the $\Delta z$ distribution around its median value. This estimator, $\sigma_{68}$, measures the width of the core of the $\Delta z$ distribution. In particular, it is defined as half of the width of the distribution, measured with respect to the median, where 68\% of the data are enclosed. 
    \item We estimate the fraction of photo-$z$ outliers by calculating the fraction of objects with 3$\sigma$ deviations in $\Delta z$, $\mathrm{out}_{3\sigma} \equiv |\Delta z| > 3\sigma_z$, where $\sigma_z$ is the standard deviation of the $\Delta z$ distribution. 
\end{enumerate}
For the photo-$z$ estimation procedure described above, we find $\sigma_{68} = 0.124$ and $\mathrm{out}_{3\sigma} = 0.017$ for the deep SN fields, and $\sigma_{68} = 0.124$ and $\mathrm{out}_{3\sigma} = 0.015$ for the shallow SN fields. For the wide-field DES survey, the photo-$z$ requirements set prior to the start of the survey specified $\sigma_{68} < 0.12$ and $\mathrm{out}_{3\sigma} < 0.015$ for 90\% of the sample of galaxies. The values we find are slightly above these requirements, but it is important to point out that galaxy samples from the SN fields reach significantly fainter magnitudes than wide-field DES galaxies, and hence it is more difficult to satisfy the wide-field requirements. However, the SOM model is constructed such that it is primarily sensitive to color, and the color-redshift relation should be agnostic to galaxy brightness. In addition, the numbers we find are similar to those reported by several photo-$z$ codes in \citet{Sanchez2014}, demonstrating the comparable performance of our method when applied to fainter galaxies. 

\subsubsection{Profile Fitting}
\label{subsec:profile-fit}
Both the measured profile, which was used to calculate DLR values and described in Section~\ref{subsec:dlr}, and the \textit{intrinsic} galaxy light profile are used in this work.

The intrinsic light profile is used to determine the location of each simulated SN within its assigned host galaxy (see Section~\ref{subsec:host-matching}). To calculate the intrinsic light profile, each galaxy in the catalog is fit with a Sérsic profile \citep{sersic} that describes the variation of galaxy intensity $I$ with radius $R$: 
\begin{equation}
\label{eq:sersic}
    I(R) = I_e \; \mathrm{exp} \Bigl\{ -b_n \Bigl[ \Bigl(\frac{R}{R_e}\Bigr)^{1/n} - 1 \Bigr] \Bigr\}
\end{equation}
$I_e$ is the galaxy intensity at the half-light radius $R_e$, $n$ is the Sérsic index describing the cuspiness of the profile, and $b_n$ is a known function of $n$. The fitting was performed by \texttt{sextractor}, which fits for $R_e$ as well as additional parameters \texttt{theta} and \texttt{aspect}, which describe the angle relative to the positive RA axis and the ellipticity of the galaxy, respectively. These additional parameters enable us to model each galaxy as an ellipse and determine its semimajor and semiminor axes, which we denote $a, b$. These quantites are calculated with $R_e = \sqrt{ab},\; \texttt{aspect} = \frac{a}{b}$. The fitted Sérsic half-light radii were scaled by a factor of 0.8 to obtain better data-simulation agreement (see Section~\ref{subsec:data-sim-agreement} for details on the data-simulation matching procedure and Section~\ref{subsec:sersic-scale} for comparisons with and without this scaling factor). This is notably the same scaling factor found to best match the DES3YR data when comparing distributions of host galaxy surface brightness at the SN position \citep[see Figure 6 of][]{des3yr}.

\subsubsection{Catalog Cuts and Parameter Fitting}
\label{subsec:catalog-cuts-params}

\begin{table}
    \centering
    \caption{Cuts applied to the host galaxy library and resulting galaxy counts.}
    \label{tbl:catalog-cuts}
    \begin{tabular}{l c}
        \hline
        Cut Requirement & Galaxies Remaining \\
        \hline
        Full catalog & 8,401,139 \\
        Removed duplicate galaxies & 7,860,305 \\ 
        Has photo-$z$ or spec-$z$ & 5,118,585 \\
        Has CIGALE galaxy parameter fit & 5,108,812 \\
        $5 < \mathrm{log}(M_{\star} / M_{\odot}) < 14$ & 4,938,372 \\
        Has reasonable Sérsic fit & 4,221,001 \\
        
        \hline
        \textbf{Final} & \textbf{4,221,001}\\
    \end{tabular}
\end{table}

Several selection criteria are applied to select \texttt{sextractor} sources that balance the trade-off between preserving realistic catalog depth/density, and maintaining the quality of the galaxy photometry. The list of cuts as well as the number of galaxies remaining after each is shown in Table~\ref{tbl:catalog-cuts}.

First, duplicate observations of galaxies are removed, prioritizing deep field observations when possible. The duplication is due to the slight overlap of certain DES SN fields, causing galaxies in the overlapping regions to appear in multiple coadded images. Next, galaxies without a spectroscopic redshift determination or photometric redshift estimate are removed, because simulated SNe cannot be associated with such galaxies. As stated in Section~\ref{subsec:hostgal-catalog}, galaxies fainter than $i=25.5$ mag do not have reliable photometric redshift estimates and are removed. In addition to the $i=25.5$ mag cut, 3.9\% of galaxies in the deep SN fields and 6.7\% of galaxies in the shallow SN fields were not successfully assigned a redshift estimate by the SOM and are removed as well.

Certain galaxy properties are known to be correlated with SN Ia rate \citep[S06;][]{mannucci,graur2017,wiseman2021}, and we use these known correlations to assign suitable host galaxies in the simulation. We estimate the total stellar mass $(M_{\star})$ and the star formation rate (SFR) for galaxies in our catalog using CIGALE \citep{cigale}. CIGALE uses grid search to find the best-fit (lowest-$\chi^2$) combination of user-specified model parameters given galaxy photometry and redshift estimates. For the galaxy parameter fits, we assume a delayed star formation history, where SFR is defined by
\begin{equation}
    \mathrm{SFR}(t) \propto \frac{t}{\tau^2} \mathrm{exp}(-t/\tau),\; 0 \leq t \leq t_0
\end{equation}
with $t_0$ the age of the onset of star formation and $\tau$ the time at which the SFR peaks. The \texttt{bc03} \citep{bc03} library of single stellar populations with a Salpeter initial mass function is used to compute the intrinsic stellar spectrum. Attenuation from dust and other sources is parameterized by the \citet{calzetti_2000} starburst attenuation curve extended with the \citet{leitherer} curve. Nebular emission is modeled by templates from \citet{inoue_2011}.
We remove a small subset of galaxies with poorly constrained CIGALE parameter fits by restricting our library to galaxies with $5 < \mathrm{log}(M_{\star}/M_{\odot}) < 14$.

We found by manual inspection that in some cases, such as very diffuse galaxies with low Sérsic index, the Sérsic profile fit fails catastrophically and produces greatly exaggerated estimates of $R_e$. This will result in SNe placed very far from the galaxy center in simulations, potentially creating a clear mismatch with SN-galaxy separations measured from the DES data. To remove galaxies with these pathological fits, we calculate ellipse areas from \texttt{sextractor} parameters ($A_{\texttt{sextractor}} = \texttt{A\_IMAGE}*\texttt{B\_IMAGE}$) as well as from Sérsic ellipse parameters ($A_{\text{Sérsic}} = ab$). We select galaxies with a ``reasonable" Sérsic fit, which we define to be $\frac{A_{\mathtt{sextractor}}}{A_{\text{Sérsic}}} \geq 0.25$.

\subsection{DES Data Host Matching}
\label{subsec:des-matching}

\begin{table}
    \centering
    \caption{Summary of host galaxy matching for the DES Y5 sample.}
    \label{tbl:des-matching}
    \begin{tabular}{l c c}
        \hline
         & Number of SNe & Percent of Total \\
        \hline
        Total sample size & 2,186 & 100\%\\ 
        Has $\geq 1$ host match & 2,047 & 94\%\\ 
        Has $\geq 2$ host matches & 126 & 6\%\\
        \hline
    \end{tabular}
\end{table}

Host galaxy matching for the DES Y5 sample is performed using the DLR method with the host galaxy catalog described in Section~\ref{subsec:hostgal-catalog}. \ddlr values are computed for all galaxies in the catalog within 15\arcsec of the SN position. Galaxies with \ddlr $> 4$ are discarded, following the SDSS convention \citep{sako2014}, and those with lowest and second lowest \ddlr values are identified as the most likely and next most likely host match (\texttt{HOSTGAL1} and \texttt{HOSTGAL2}). If fewer than two galaxies have \ddlr $\leq 4$, those SNe are considered to be missing a host match or (if one galaxy has \ddlr $\leq 4$) missing a second host match. Summary statistics for the DES data with host matches is shown in Table~\ref{tbl:des-matching}.

Ideally, matching should be done with the full catalog, including galaxies without a known redshift. This would allow for us to remove SNe hosted by galaxies without known redshifts, rather than match them incorrectly to a galaxy with known redshift. However, we find that $\sim 98\%$ of DES SNe have the same host match when matched with the full catalog, and those that do not mostly become ``hostless" when the cuts are employed. Thus, we conclude that matching with the cut catalog is a valid method, as these SNe are ineligible for the analysis regardless of the cuts. 

\section{Simulations and event selection}
\label{sec:sims}

\subsection{Simulations}
All simulations for this work are produced with the SuperNova ANAlysis (SNANA) software \citep{snana}. The SNANA simulation starts with a SN spectral template and generates survey-specific photometry under realistic observing conditions by utilizing a cadence library containing zero points, sky noise, and PSF information for each telescope pointing on each observing night.

We note that overlaying simulated SNe on actual galaxy images is the ideal method of performing a host matching analysis; however, it would be very computationally expensive to run the volume of simulations needed to develop and constrain the host mismatch systematic. For improved computational efficiency, we use the catalog-level simulation from SNANA.

\subsubsection{SN Models}
\label{subsec:sn-models}

SN Ia simulations for this work are created using the SALT2 model \citep{salt2} with training parameters determined from the Joint Lightcurve Analysis \citep[JLA,][]{jla}. The SALT2 model defines several restframe parameters for SN lightcurves: the time of SN peak brightness $t_0$, a stretch-like parameter $x_1$, a color parameter $c$ and the lightcurve normalization parameter $x_0$. Nuisance parameters $\alpha, \beta$ are determined according to \citet{scolnic2016}, while the color ($c$) and stretch ($x_1$) populations follow \citet{popovic2021}. The SNe Ia are simulated with a redshift-dependent volumetric rate, using measured rates from \citet{dilday} and \citet{perrett} and recomputed by \cite{frohmaier}. To model empirically measured Hubble scatter after the SN Ia standardization procedure, we use the spectral-variation intrinsic scatter model in \citet{k13} that is based on the model uncertainties determined in \cite{g10}. The simulations include two separate populations: the DES-like photometric sample and a spectroscopically confirmed low-$z$ anchor. For simplicity, the low-$z$ anchor is an ad-hoc DES simulation applied to $0 < z < 0.1$ with an inflated rate to match the true number of low-$z$ events in the DES-SN5YR sample.

Two types of core-collapse SNe as well as two types of peculiar SNe Ia are simulated alongside the SNe Ia to evaluate the effects of host galaxy mismatch on redshift-dependent photometric classification: SNII, SNIbc, SNIax, and SNIa-91bg. These simulations use the SED templates introduced in \cite{v19} with luminosity functions and rates following \cite{vincenzi2021}. We use relative rates as measured by \citet{shivvers} anchored by an overall rate following the cosmic star formation history presented in \cite{madau} normalized by the local SN rate from \cite{frohmaier20}.

All simulations are generated assuming a flat $\Lambda$CDM cosmology with $H_0 = 70\; \mathrm{km\;s^{-1}\;Mpc^{-1}}$ and $\Omega_{\mathrm{M}} =0.311$.

\begin{figure}
    \includegraphics[scale=0.38,trim={0cm 0cm 0cm 0cm}]{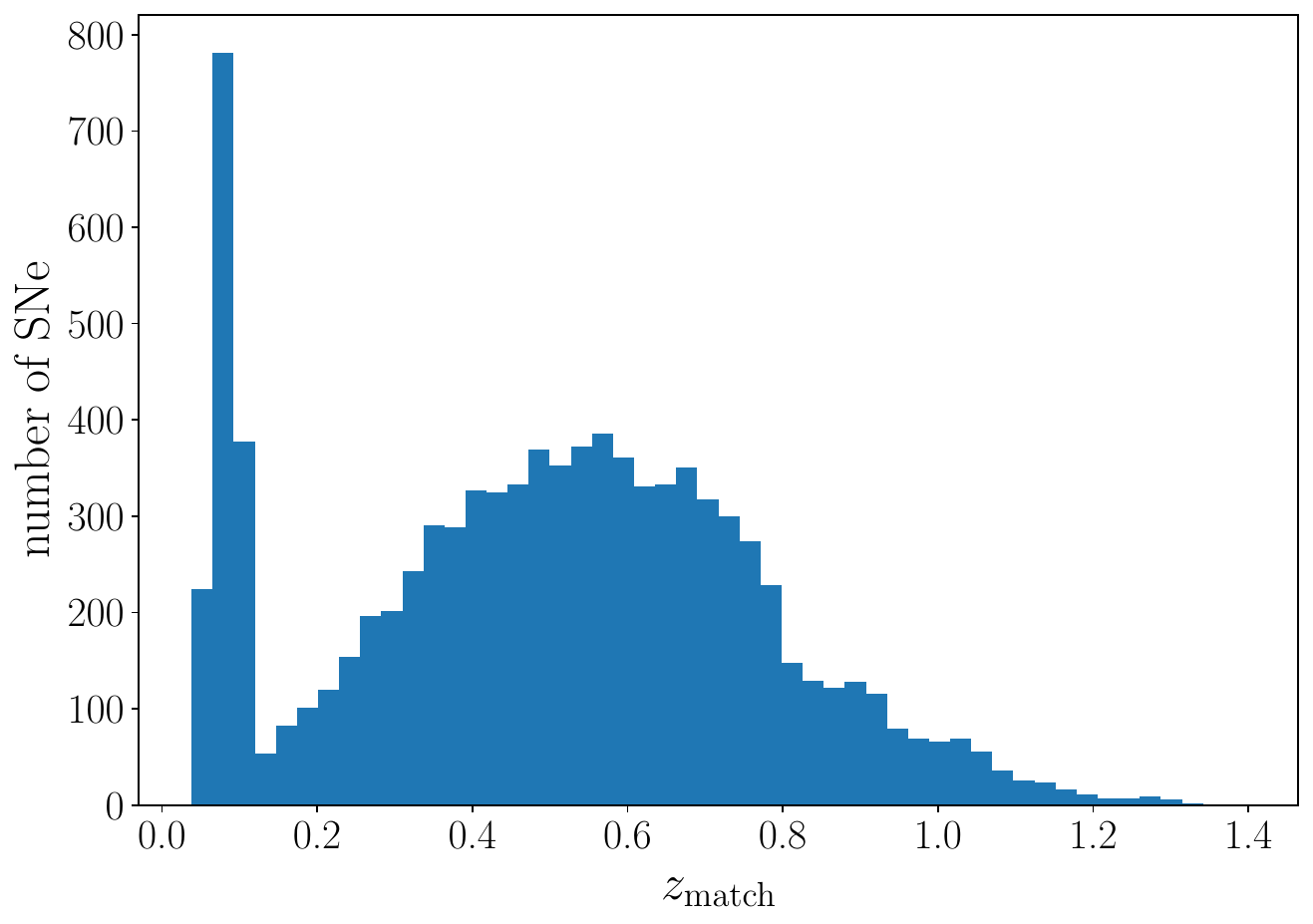}
    \centering
    \caption{Redshift distribution of matched host galaxies in one realization of the SNIa-only simulations. Since we find the mismatch rate resulting from the DLR method to be quite low (see Table~\ref{tbl:sim-cuts-Ia-cc}), the true redshift distribution is not visibly different and was not included in this plot.}
    \label{fig:Ia_z_dist}
\end{figure}

\begin{figure*}
    \includegraphics[scale=0.38,trim={0cm 0cm 0cm 0cm}]{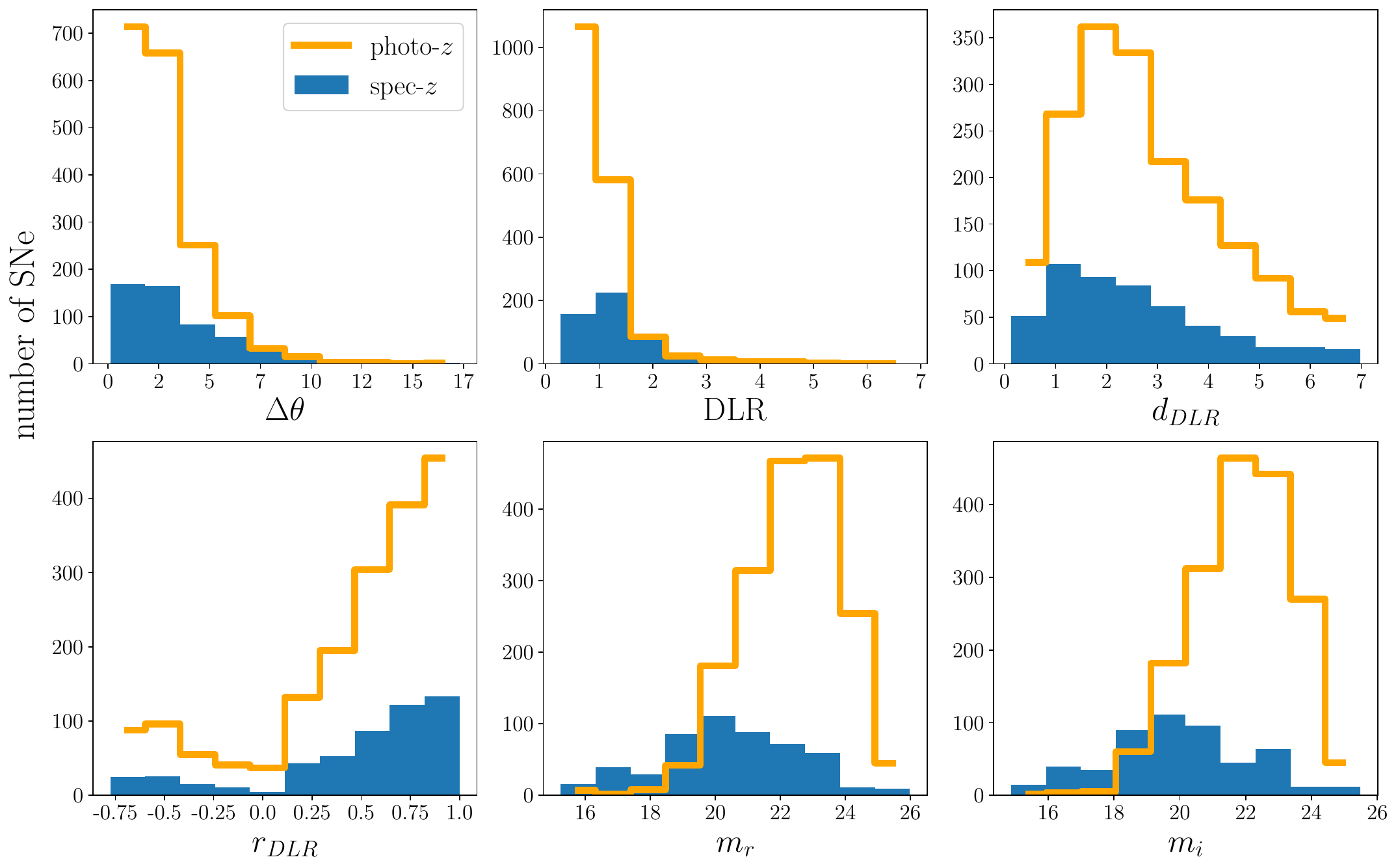}
    \centering
    \caption{Parameter distributions of matched host galaxies in one realization of the SNIa-only simulations split into two populations: galaxies with spectroscopic redshifts and those with photometric redshift estimates only. Details and definitions of these parameters can be found in Section~\ref{subsec:data-sim-agreement}. Since we find the mismatch rate resulting from the DLR method to be quite low (see Table~\ref{tbl:sim-cuts-Ia-cc}), the true distributions are not visibly different and were not included in this plot.}
    \label{fig:spec_vs_phot_dists}
\end{figure*}

\subsubsection{Host Association and Matching}
\label{subsec:host-matching}
To associate a simulated SN with a host galaxy, the simulation first generates properties for each SN, including color, stretch, redshift, and sky location (RA, DEC). Subsequently, all galaxies whose redshifts match the true SN redshift within a small tolerance ($dz_{\mathrm{tol}}= \text{max} |z_{\mathrm{SN}} - z_{\mathrm{GAL}}| = 0.002 + 0.04z$) are selected. The \textit{assigned} host galaxy is chosen from this subset using the host mass-dependent weighting shown in \cite{wiseman2021}. The simulated host association and matching use the same deep coadded DES galaxy catalog used for the data analysis.

Note that the process of assigning a host for a simulated SN does not take into account the locations of the host or the SN, only the redshifts of the SN and galaxy as well as the galaxy stellar mass. The host and its neighboring galaxies are moved near the SN such that it satisfies the simulated SN-host separation as well as models host confusion in the analysis. This strategy is efficient for modeling SN-host correlation and incorrect host matches, but it does not model large scale structure.

The SN-host separation is determined by placing the SN at a radial distance $R$ from the center of the assigned host galaxy according to the probability distribution $p(R) \sim I(R)$, where $I(R)$ is the galaxy intensity at radius $R$ described by the galaxy's fitted Sérsic profile (see Equation~\ref{eq:sersic}). This approach follows past work using SNANA such as \citet{des3yr}.

To replicate the host matching procedure used for real data, we apply host galaxy matching with the DLR method to the simulated SNe. The simulation computes $d_{\mathrm{DLR}}$ values for up to 10 galaxies within a 10\arcsec\ radius of the assigned host and saves the galaxies with the lowest and second lowest $d_{\mathrm{DLR}}$ values as the most and next most likely host match (\texttt{HOSTGAL1} and \texttt{HOSTGAL2}), as long as $d_{\mathrm{DLR}}\leq 4$. The redshift distribution of the matched host galaxies for one realization of a SNIa-only simulation is shown in Figure~\ref{fig:Ia_z_dist} and relevant population parameter distributions (e.g. $d_{\mathrm{DLR}}$) are shown for host galaxies with spectroscopic and photometric redshifts in Figure~\ref{fig:spec_vs_phot_dists}.
The host matching procedure is run for the full set of simulations, including the low-$z$ population, though low-$z$ mismatches are rare. 

\subsection{Event Selection}
\subsubsection{Lightcurve Fitting}
\label{subsec:lcfit}

All DES-SN observed and simulated SN lightcurves are fit with the same SALT2 model with JLA parameters that was used to simulate the SNe Ia. The fit is performed with a $\chi^2$-minimization program included in SNANA and determines several parameters under the assumption that the event is a SN Ia: the time of SN peak brightness $t_0$, a stretch-like parameter $x_1$, a color parameter $c$ and the lightcurve normalization parameter $x_0$, as well as their uncertainties and covariances (i.e., $\sigma_{t_0}$, etc.). These parameters are used to calculate the distance modulus $\mu$, allowing the SNe to be placed on the Hubble diagram.

We apply selection cuts on these fitted parameters and select SN lightcurves well described by the SALT2 model. Specifically, we restrict our sample to SNe satisfying the following criteria:

\begin{itemize}
  \item $|x_1| < 3$,
  \item $|c| < 0.3$,
  \item $\sigma_{x_1} < 1$, and
  \item $\sigma_{t_0} < 2$ days.
\end{itemize}

\begin{figure*}
    \includegraphics[scale=0.3,trim={0cm 0cm 0cm 0cm}]{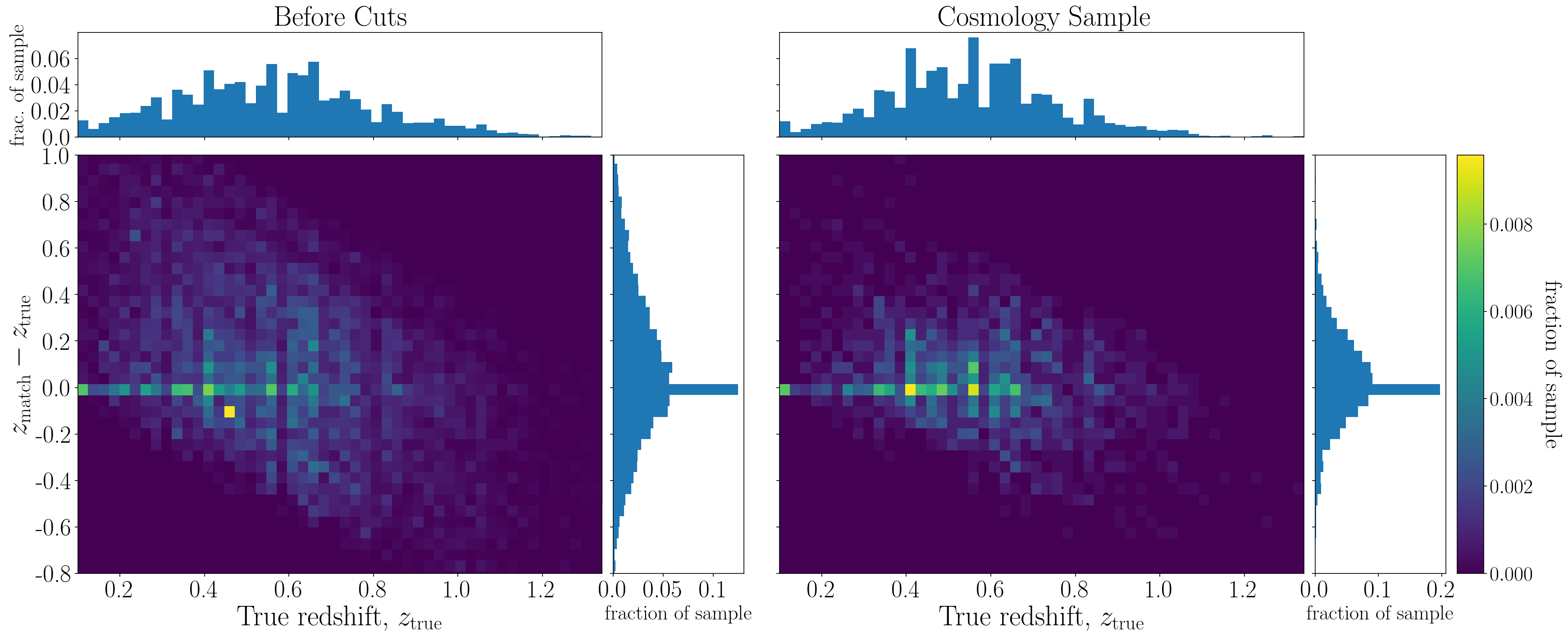}
    \centering
    \caption{Redshift differences ($\Delta z \equiv z_{\mathrm{match}} - z_{\mathrm{true}}$) for simulated SNe Ia with mismatched hosts. (left) $\Delta z$ as a function of $z_{\mathrm{true}}$ for the full simulated sample prior to applying any selection cuts. (right) $\Delta z$ as a function of $z_{\mathrm{true}}$ after all selection cuts (described in Section~\ref{subsec:lcfit}) and removal of SNe without a valid bias correction (described in Section~\ref{subsec:bcor}). Our selection process not only reduces the mismatch rate (see Table~\ref{tbl:sim-cuts-Ia-cc}) but also removes SNe with extremely biased redshift estimates from mismatched hosts, reducing the spread in $\Delta z$. Quantitatively, the spread in $z_{\mathrm{match}} - z_{\mathrm{true}}$ characterized by the middle 90\% of the distribution is reduced from 1.1 before cuts to 0.6 after cuts.}
    \label{fig:Ia_delta_z}
\end{figure*}

A summary of these cuts on our simulations can be found in Table~\ref{tbl:sim-cuts-Ia-cc}. The right panel of Figure~\ref{fig:Ia_delta_z} shows the effect of these cuts on the redshift error distribution between the matched and true hosts. These data show that the cuts not only reduce the average mismatch rate over 25 realizations from 2.5\% to 1.7\%, but also significantly reduce the spread in redshift error. Quantitatively, the cuts reduced the middle 90\% of the $z_{\text{match}}-z_{\text{true}}$ distribution for mismatched pairs from 1.1 (left panel of Figure~\ref{fig:Ia_delta_z}) to 0.6 after cuts (right panel).

\begin{table*}
    \centering
    \caption{Summary of the mismatch rate averaged over 25 realizations after each selection cut on the SNIa+CC simulated dataset, split by Ia vs. non-Ia SNe. The non-Ia SNe include SNIax, SNIa-91bg, SNII, and SNIbc. All results using SNe Ia only are using the SN Ia subset of this dataset.}
    \label{tbl:sim-cuts-Ia-cc}
    \begin{tabular}{l c c c c c c}
        \toprule
        Cut & \multicolumn{3}{c}{SN Ia} & \multicolumn{3}{c}{non-Ia SNe}\\
        \cmidrule(lr){2-4} \cmidrule(lr){5-7} 
         & Mismatches & Total & Mismatch Rate & Mismatches & Total & Mismatch Rate\\
        \midrule
        No cuts & 233 & 9,356 & 2.5\% & 151 & 5,957 & 2.5\%\\ 
        $|x_1| < 3$, $|c| < 0.3$ & 95 & 5,562 & 1.7\% & 22 & 800 & 2.8\%\\ 
        $\sigma_{x_1} < 1$ & 83 & 4,886 & 1.7\% & 15 & 545 & 2.8\%\\
        $\sigma_{t_0} < 2$ & 83 & 4,870 & 1.7\% & 15 & 543 & 2.8\%\\
        \bottomrule
    \end{tabular}
\end{table*}

\subsubsection{Photometric Classification}
\label{subsec:classification}

In the absence of SN spectra, we rely on photometric classifiers to remove non-Ia contaminants from the cosmological sample. We choose two recent neural-network-based classifiers with high demonstrated accuracies for this work: SuperNNova \citep[SNN,][]{supernnova} and SCONE \citep{scone}. High quality SN redshifts have been shown to improve SNN accuracy classifying SN Ia vs. non-Ia, so we test SNN in both redshift-dependent and redshift-independent configurations to evaluate the impact of misidentified redshifts. SNN models were trained in both the redshift-independent and dependent configurations following \citet{Vincenzi_2022}. SCONE is redshift-independent and performs classification based on SN lightcurves alone, so these results should not be affected by host misidentification. Each classifier outputs $P_{\mathrm{Ia}}$ values, the predicted probability of each SN to be a type Ia. The SNIa-only simulations are not run through photometric classification; all objects are simply labeled as SNe Ia.

\section{Comparing Data with Simulations}
\label{subsec:params}

\subsection{Host Matching Rates and Hostless SNe}

First, we compare the fraction of simulated and observed DES SNe  that pass our selection cuts with one or more matched hosts. Since we have reliable SN Ia photometric classifiers, we compare simulated true SNe Ia with DES SNe Ia as predicted by SCONE \citep{scone}. Since most of the redshifts for the DES SNe come from host galaxies, we chose to use SCONE for this comparison as it does not require redshift information.

After applying the cuts specified in Table~\ref{tbl:sim-cuts-Ia-cc} to both data and simulations, the DLR algorithm is able to find at least one host galaxy match in 98.3\% of SNe in our SN Ia-only simulations compared to 98.2\% of SCONE-classified SNe Ia in the DES data. The remaining $\sim 2$\% of SNe in our data and simulations is the rate of ``hostless" SNe. SNe can appear hostless because their host galaxies are too faint to be detected and our imposed \ddlr $\leq 4$ cut on potential host matches avoids matching to unrealistically faraway galaxies.

SNe in galaxy-dense regions have multiple potential host matches with \ddlr $\leq 4$. We find that ($9.35\pm 0.09$)\% of simulated SN Ia have more than one host match, compared to ($8.71\pm 1.06$)\% of SCONE-classified DES SNe Ia. The agreement in the fraction of hostless and multiple-host events
provides confidence in our simulated data sample.

\begin{figure*}
    \includegraphics[scale=0.35,trim={0cm 0cm 0cm 0cm}]{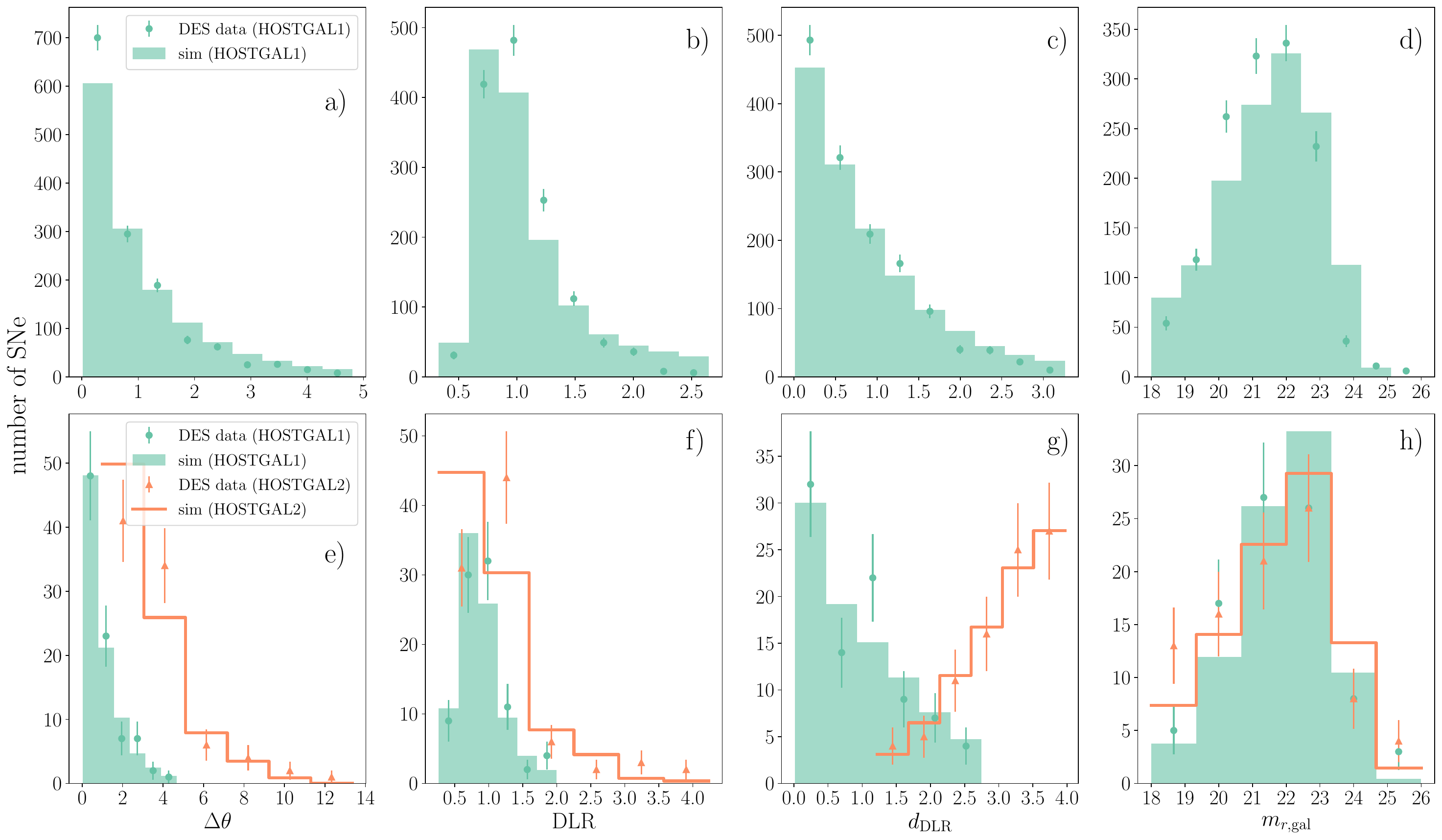}
    \centering
    \caption{Histograms comparing our simulated SNe Ia against DES Y5 photometrically confirmed SNe Ia for angular SN-galaxy separation ($\Delta \theta$, arcsec), directional light radius (DLR), the DLR-normalized SN-galaxy separation ($d_{\mathrm{DLR}} = \frac{\Delta \theta}{\mathrm{DLR}}$), and $r$-band host galaxy magnitude ($m_{r,\mathrm{gal}}$) (see Section~\ref{subsec:params} for explanations of each). For ease of comparison, the simulation histograms are normalized to match the integral of the data histogram and the $x$ axis limits of each histogram are determined by the middle 90\% of the data distribution to remove outliers. In both rows, points with error bars represent parameter distributions measured from the DES data and filled or unfilled histograms represent the analogous quantities for simulations. (top row) Histograms of parameter values for closest host galaxy match (\texttt{HOSTGAL1}) for SNe with only one host galaxy match. (bottom row) Histograms of parameter values for closest (\texttt{HOSTGAL1}) and second closest (\texttt{HOSTGAL2}) host galaxy match for SNe with 2 or more host galaxy matches.}
    \label{fig:param-dists}
\end{figure*}

\subsection{Comparing Parameter Distributions}
\label{subsec:data-sim-agreement}

To further verify that the host mismatch rate estimates and resulting cosmological biases derived from our simulations are representative of the DES sample, we compare simulations and data over five relevant parameter distributions: SN-galaxy separation ($\Delta \theta$), DLR, $d_{\mathrm{DLR}}$, $r$-band host galaxy magnitude ($m_{r, \mathrm{gal}}$), and the \texttt{HOSTGAL1}  to \texttt{HOSTGAL2}  \ddlr ratio ($r_{\mathrm{DLR}}$), following \cite{popovic2020}. A comparison of these distributions for the DES-SNY5YR sample (shown in points) and the simulations (shown in filled/unfilled bars) used for this analysis is shown in Figure~\ref{fig:param-dists}.

The top row of Figure~\ref{fig:param-dists} shows the parameter distributions of data and simulations for SNe with a single host galaxy match (i.e. only one galaxy with \ddlr $\leq 4$). The second row shows the same parameter distributions for SNe with at least two host galaxy matches, where green show distributions for the closest host galaxy match (smallest \ddlr value, labeled \texttt{HOSTGAL1}) and orange denotes the second closest host galaxy match (labeled \texttt{HOSTGAL2}).

The angular separation between the center of the galaxy and the SN position, {$\Delta \theta$, is shown in panels a) and e) of Figure~\ref{fig:param-dists}. Good agreement in the data vs. simulation $\Delta \theta$ distribution validates that (1) the algorithm used by the simulation to place SNe within their host galaxies is representative of real observations, and (2) the Sérsic ellipse parameters $a,b$ used to place SNe within their host galaxies are well estimated. Note that, as described in Section~\ref{subsec:profile-fit}, the fitted Sérsic parameters were scaled by a factor of 0.8. The distributions from data and simulations match very well overall, but the simulations slightly underestimate \texttt{HOSTGAL1} matches at the very low end of the $\Delta \theta$ distribution.

DLR, shown in panels b) and f), corresponds to the size of the matched host and is measured as in Section~\ref{subsec:dlr}. Agreement in this parameter verifies that matched hosts are similar in size between data and simulations. The data and simulations agree well for both \texttt{HOSTGAL1} and \texttt{HOSTGAL2}.

The distribution of \ddlr values is shown in panels c) and g). \ddlr agreement is an important quantity, since it is used to determine which galaxies are host matches, and shows that simulated SNe are placed at reasonable distances from the host center. These distributions largely show the same trend as the $\Delta \theta$ distributions, where the \texttt{HOSTGAL2} distributions match very well but the \texttt{HOSTGAL1} distributions from simulations appear to be skewed slightly higher than those of the DES data.

$m_{r,\text{gal}}$, the host galaxy $r$-band magnitude, is shown in panels d) and h). Agreement in this parameter is an additional validation of similarity in the overall populations of host galaxies between data and simulations. The slight discrepancy between these distributions can likely be attributed to the fact that the spectroscopic efficiency is defined using \texttt{MAG\_AUTO}, but we applied the efficiency to \texttt{MAG\_MODEL} magnitudes, which tend to be slightly fainter.

Finally, we define
\begin{equation}
r_{\mathrm{DLR}} = \frac{d_{\mathrm{DLR, HOSTGAL1}}}{d_{\mathrm{DLR, HOSTGAL2}}}.
\end{equation}
Figure~\ref{fig:rddlr_cigale} shows the distributions of this parameter for DES data and our simulations. Since the definition requires $d_{\mathrm{DLR, HOSTGAL2}}$, only SNe with at least 2 host galaxy matches are included. Consistency in $r_{\mathrm{DLR}}$ and the \texttt{HOSTGAL2} distributions indicate that the galaxy catalog is sufficiently dense, since the spacing between galaxies will affect the \texttt{HOSTGAL2} parameter distributions much more than \texttt{HOSTGAL1}.

\begin{figure}
    \includegraphics[scale=0.5]{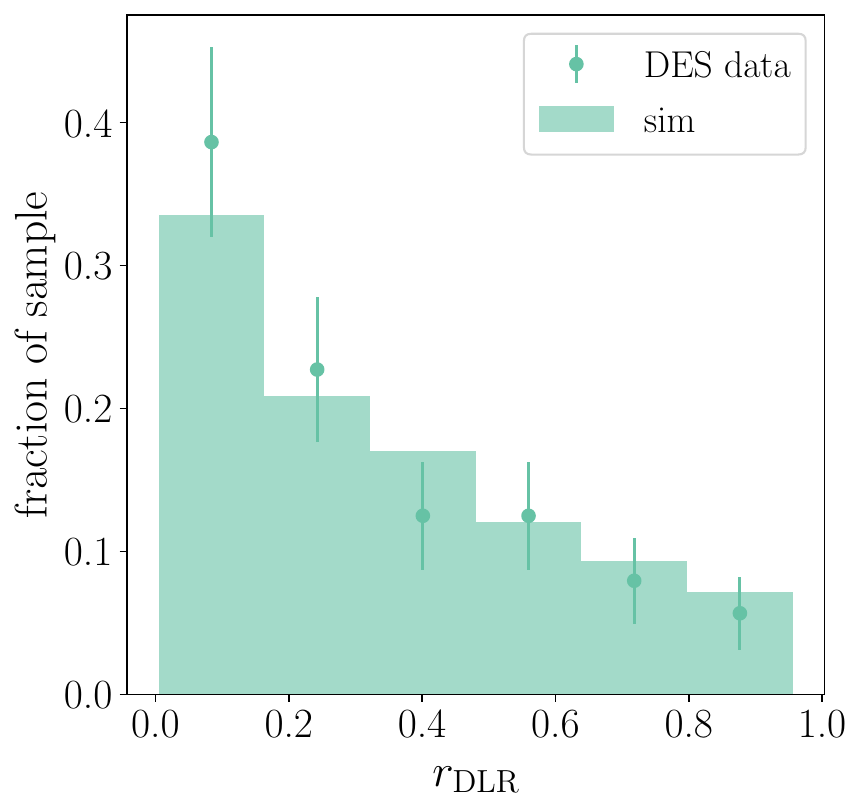}
    \caption{Distributions of the ratio $r_{\mathrm{DLR}} = \frac{d_{\mathrm{DLR, HOSTGAL1}}}{d_{\mathrm{DLR, HOSTGAL2}}}$} for DES Y5 photometrically confirmed SNe Ia and our simulated SNe Ia.
    \label{fig:rddlr_cigale}
\end{figure}


\section{Cosmological Parameter Estimation}
\label{sec:cosmo}

\subsection{Cosmology Analysis Overview}
The cosmology analysis pipeline is orchestrated end-to-end by the \texttt{Pippin} framework \citep{pippin}, beginning with simulations or data as input and concluding with cosmological parameter estimates. After SNe are simulated, the resulting lightcurves are fit with an empirical model that outputs characteristics for each SN, such as color and stretch (see Section~\ref{subsec:lcfit}). SNe that are fit successfully are then assigned a SN Ia probability by a photometric classifier (see Section~\ref{subsec:classification}) and bias corrections on distance moduli are computed for each SN based on its fitted parameters as well as SN Ia probability. Finally, the bias corrected distance moduli are used to determine constraints on cosmological parameters.

\subsection{Bias Corrections}
\label{subsec:bcor}

Biases due to core collapse contamination and survey selection are modeled and corrected for with the BEAMS with Bias Corrections framework \citep[BBC,][]{bbc}, an extension of BEAMS \citep{beams}, which allows photometrically classified SNe Ia to be used for cosmology. The primary output of the BBC framework applied on a SN sample is a redshift-binned Hubble diagram corrected for biases from selection effects and non-SNIa contamination.

First, systematic biases due to selection effects are modeled by a large simulation of SNe Ia ($\sim 800,000$ SNe Ia). We found empirically that our current framework for modeling and correcting for biases, typically used for e.g. the Malmquist bias, is not suited for including wrong hosts in our bias correction simulations. Thus, our bias correction simulation includes only assigned hosts for our primary results. In Section~\ref{subsec:bcor-mismatch}, we explore incorporating host matching (and thus mismatched hosts) into the bias correction simulations. 

Using this large SN sample, distance moduli are calculated using the Tripp formula \citep{tripp}, 
\begin{equation}
\label{eq:tripp}
\mu_\mathrm{obs} = m_B + \alpha x_1 - \beta c + M_B+\Delta \mu_{\mathrm{bias}}
\end{equation}
for each SN. $m_B=-2.5 \mathrm{log_{10}}(x_0)$ and $M_B$ is the absolute magnitude of a SN Ia with $x_1 = c = 0$ and $\alpha,\beta$ are nuisance parameters determined according to \citet{scolnic2016}. Biases from a reference cosmology, $\Delta \mu_{\mathrm{bias}}$, are calculated in a 3-dimensional grid of $\{z, x_1, c\}$ bins using the method described above. We use this grid of estimated biases to correct all of our distance moduli prior to cosmology fitting. A small percentage of SNe with parameter values that do not fit into the grid are discarded. 
Finally, the BEAMS method is used to estimate binned distance moduli from the bias-corrected distance moduli from the previous step in the presence of core-collapse contamination. This is done by minimizing the BEAMS likelihood, which models the SNe Ia population and a population of contaminants separately. These terms are weighted by $P_{\mathrm{Ia}}$, the probability of each SN to be a type Ia as output by a photometric classifier. We omit the mass step correction in Equation~\ref{eq:tripp}; details on the mass step and the impact of host mismatches can be found in Section~\ref{subsec:mass-step}.

\subsection{Cosmological Parameters}
\label{subsec:cosmo-params}
 We fit for $w$ and $\Omega_m$ using \texttt{wfit}, a fast cosmology grid-search program in SNANA,  assuming a diagonal covariance matrix $\mathcal{C}_{\text{stat}}$ and an approximate CMB prior computed with the $R$-shift parameter (see e.g. Equation 69 in \citet{WMAP:2008lyn}) from the same cosmological parameters used to generate the SNe Ia. The $R$ uncertainty is  $\sigma_R = 0.006$, tuned to have the same constraining power as \citet{planck}. As we are only interested in the impact of host mismatches on cosmology, the approximation of a diagonal covariance matrix is sufficient for our purposes.
\texttt{wfit} calculates the $\chi^2$ of the SN likelihood to compute our final cosmological fit,
\begin{equation}
\label{eqn:chi2}
    \chi^2 = \Delta \mu_{\text{model}}^T \cdot \mathcal{C}_{\text{stat}}^{-1}\cdot\Delta \mu_{\text{model}}
\end{equation}
where 
\begin{equation}
    \Delta \mu_{\text{model}} = \mu - \mu_{\text{model}}(\Omega_m, w).
\end{equation}

\section{Results and Discussion}
\label{sec:results}

We use the cosmological parameter estimation framework described in Section~\ref{sec:cosmo} to evaluate the effects of host galaxy mismatch on the resulting best fit cosmology. We focus primarily on shifts in the best fit value for $w$, the dark energy equation of state parameter. We quantify this shift by creating two sets of identical simulations that differ only in whether or not the DLR method is run to determine the matched hosts using the procedure described in Section~\ref{subsec:host-matching}. We define $S_{\mathrm{match}}$ as simulations with matched hosts and $S_{\mathrm{truehost}}$ as simulations with perfect host matching. In $S_{\mathrm{truehost}}$, we do not calculate matched hosts using the DLR method; we instead force a match to the true hosts assigned by the simulation. This ensures that there will be no mismatches and serves as a baseline for comparison.

We define the $w$ shift as the difference between the inferred $w$ values from $S_{\mathrm{match}}$ ($w_{\text{match}}$) and $S_{\mathrm{truehost}}$ ($w_{\text{truehost}}$). We average over 25 realizations with the same simulation parameters. Explicitly, we define the $w$ shift as
\begin{equation}
    \Delta w = \langle w_{\textrm{match}} - w_{\textrm{truehost}} \rangle_{\mathrm{(25\;realizations)}}
\end{equation}
where $\langle \rangle$ denotes the inverse-variance weighted average. 
We calculate the associated uncertainty on $\Delta w$ as follows:
\begin{equation}
\sigma_{\Delta w} = \sqrt{\frac{\sum_i (w_{\textrm{match},i} - w_{\textrm{truehost}, i})^2}{25}}.
\end{equation}

\subsection{Cosmological Biases with SNe Ia Only}
\label{subsec:cosmo-Ia}

For simulations with SNe Ia only, we find $\Delta w = 0.0013 \pm 0.0026$. \textbf{In the flat $w_0 w_a$ CDM model, we find biases of $\Delta w_0=0.037 \pm 0.041, \Delta w_a = -0.22 \pm 0.25$. Biases in both cosmological models are consistent with zero.} In Figure~\ref{fig:Ia_delta-mu}, we show the biases on the binned Hubble diagram comparing distance moduli $\mu$ from simulations with DLR matched hosts ($\mu_{\text{match}}$) and true hosts ($\mu_{\text{truehost}}$). We define 
\begin{equation}
\label{eq:delta-mu}
    \Delta \mu = \mu_{\textrm{match}} - \mu_{\textrm{truehost}}.
\end{equation}
We see that the bias is consistent with 0 until $z \sim 1$, where the sample becomes very sparse (see Figure~\ref{fig:Ia_z_dist} for the redshift distribution of the sample).

Figure~\ref{fig:Ia_HD} shows the Hubble diagram and Hubble residuals for a single realization of simulations with mismatches. In this particular realization, 79 SNe were matched to the wrong host out of 5,811 total simulated SNe, which translates to a 1.4\% mismatch rate. The mismatched SNe, shown in red circles, show similar Hubble residuals compared to the correctly matched SNe, which is consistent with the small recovered bias on $w$. The observed similarity in Hubble residuals is likely due to the fact that catastrophic outliers in redshift are removed by the selection criteria described in Table~\ref{tbl:sim-cuts-Ia-cc} and shown in Figure~\ref{fig:Ia_delta_z}.

The binned Hubble residuals for all 25 simulations with mismatch are shown in Figure~\ref{fig:Ia_residuals}. This plot shows $\Delta \mu = \mu_{\mathrm{match}} - \mu_{\mathrm{model}}$, as opposed to Figure~\ref{fig:Ia_delta-mu}, which shows $\mu_{\mathrm{match}} - \mu_{\mathrm{truehost}}$. This allows us to compare residuals from the subpopulations of SNe with wrong and correct host match with respect to a fiducial cosmology. Aggregated over all 25 simulations, the bias from SNe with the wrong host match (shown in orange) is clearly distinct from the nearly unbiased subset of SNe with the correct host match (teal). Although wrong hosts clearly lead to biases on the Hubble diagram, SNe with the wrong host match make up $<2\%$ of the sample, resulting in the small $\Delta w$ value we observe.

\begin{figure}
    \includegraphics[scale=0.45,trim={0cm 0cm 0cm 0cm}]{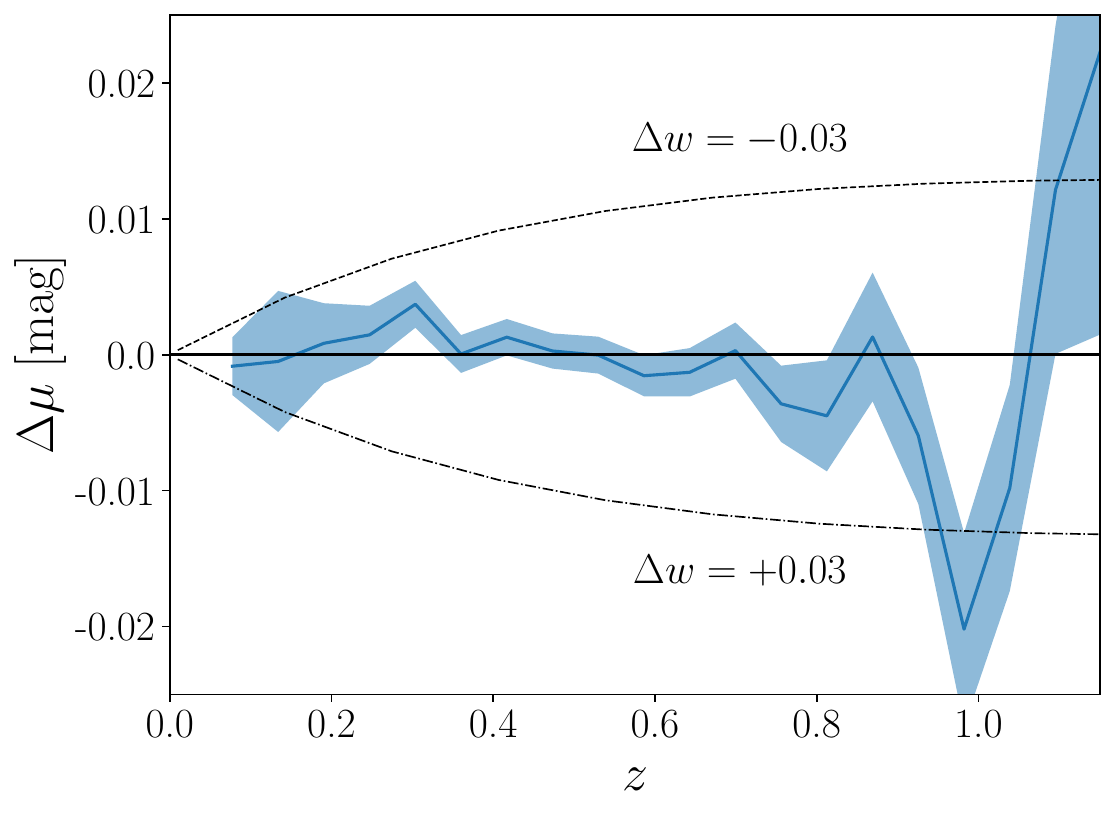}
    \centering
    \caption{Biases in binned Hubble residuals for the SN Ia-only sample between samples with and without mismatched host galaxies, $\Delta \mu = \mu_{\text{match}} - \mu_{\text{truehost}}$, as a function of redshift. Uncertainties are shown as the shaded region and calculated from the binned standard deviations of $\mu_{\text{match}}$ and $\mu_{\text{truehost}}$. Lines showing $\Delta w = \pm 0.03$ are also plotted for reference.}
    \label{fig:Ia_delta-mu}
\end{figure}

\begin{figure}
    \includegraphics[scale=0.37,trim={0cm 0cm 0cm 0cm}]{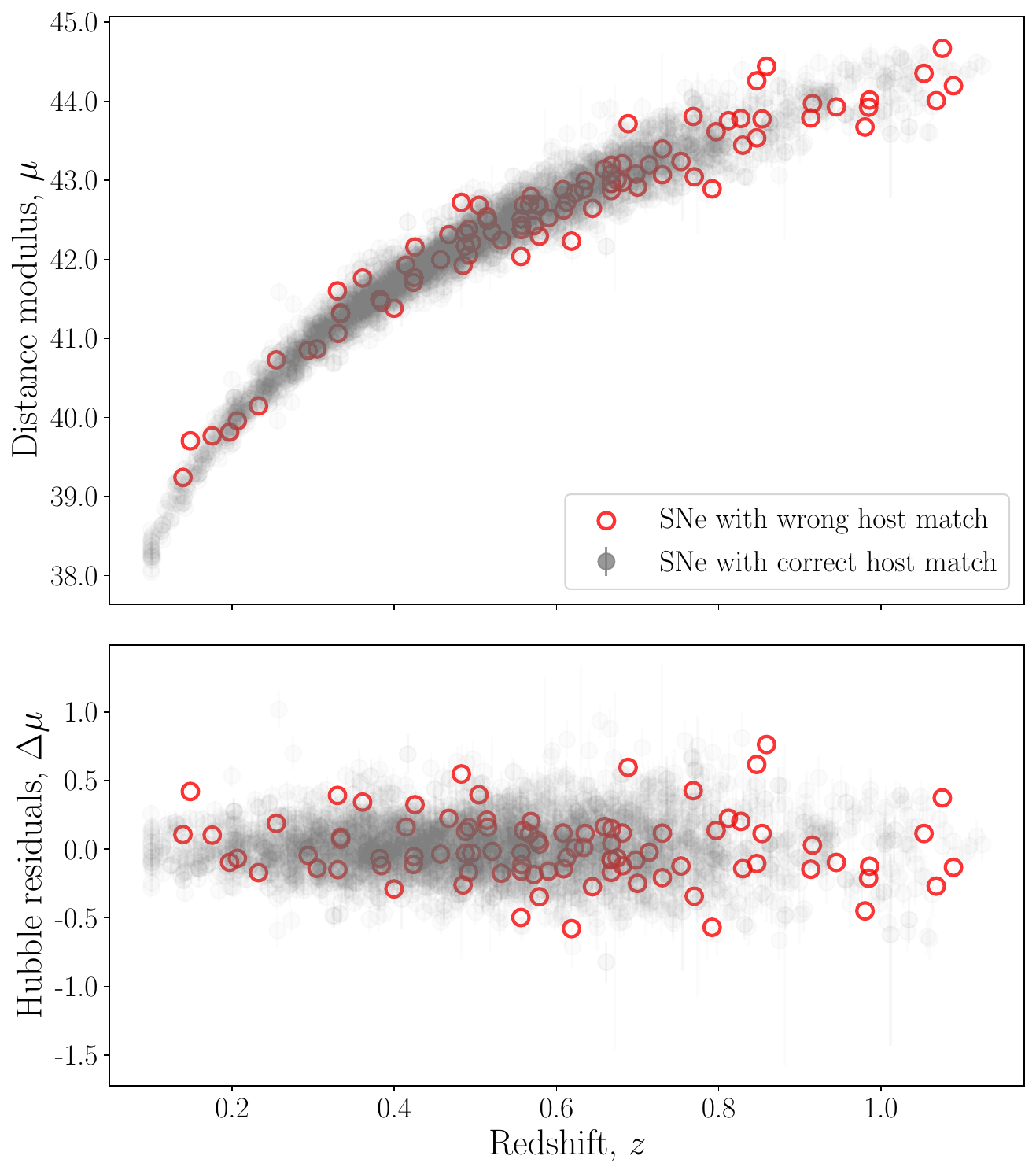}
    \centering
    \caption{Hubble diagram for a single realization of our SNIa-only simulation with matched hosts. This realization has 79 SNe matched to an incorrect host out of 5,811 total SNe, a 1.4\% mismatch rate. The low-$z$ sample is omitted from this plot, since it is not part of the main DES sample.}
    \label{fig:Ia_HD}
\end{figure}

\begin{figure}
    \includegraphics[scale=0.45,trim={0cm 0cm 0cm 0cm}]{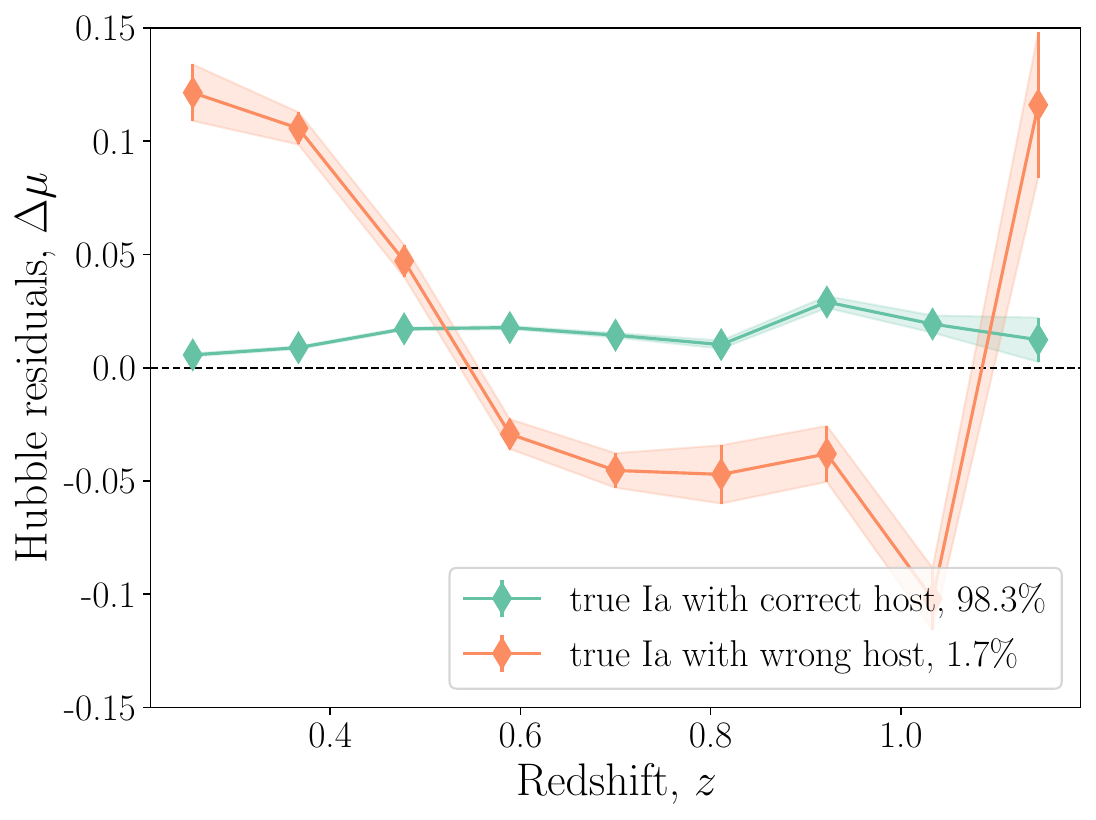}
    \centering
    \caption{Binned Hubble residuals ($\mu_{\text{match}}-\mu_{\text{model}}$) for all 25 realizations of our SNIa-only simulation with mismatch. The percentages in the legend show the fraction of the sample represented by each population in the plot.}
    \label{fig:Ia_residuals}
\end{figure}


\begin{table*}
    \centering
    \begin{tabular}{l l c c c c}
        \toprule
        Classifier & SNe & $\Delta w$ & \multicolumn{3}{c}{Ia vs. non-Ia Classification Accuracy} \\
        \cmidrule(lr){4-6} 
         & & & with host mismatch & no host mismatch & change\\
        \midrule
        Perfect & Ia only & $0.0013 \pm 0.0026$ & 100\% & 100\% & 0\% \\ 
        Perfect & Ia+CC & $0.0011 \pm 0.0027$ & 100\% & 100\% & 0\% \\ 
        \snnz & Ia+CC & $0.0032 \pm 0.0040$ & 97.94\% & 98.06\% & -0.1\% \\ 
        \snnnoz & Ia+CC & $0.0009 \pm 0.0028$ & 97.05\% & 97.05\% & 0\% \\
        SCONE & Ia+CC & $0.0016 \pm 0.0032$ & 96.13\% & 96.13\% & 0\%\\
        \bottomrule
    \end{tabular}
    \caption{$\Delta w$ and classification accuracies for each classifier. Accuracy change is only expected for the \snnz\ classifier, as it is the only classifier tested that requires redshift information. Accuracy change is defined as $\textrm{Accuracy (with mismatch)} - \textrm{Accuracy (no mismatch)}$.}
    \label{tbl:Ia-cc-results}
\end{table*}

\subsubsection{Host Mismatch and the Mass Step}
\label{subsec:mass-step}

The `mass step' is the observed correlation between SNe Ia intrinsic luminosity and host galaxy stellar mass, $M_{\star}$. Specifically, SNe Ia in more massive galaxies are more luminous after lightcurve corrections than their counterparts occurring in galaxies with lower stellar mass, with the average corrected luminosity distribution following a two-part step function with a break at log($M_{\star} / M_{\odot}) \sim 10$ \citep{kelly2010, mass_step, jla, smith2020, kelsey2021, kelsey2023}. Though the underlying astrophysical cause is unknown, recent cosmological analyses have incorporated a correction in which SN luminosities in hosts with log($M_{\star} / M_{\odot}$) $\geq 10$ and those in hosts with log($M_{\star} / M_{\odot}$) $< 10$ are fit for separately. When this two-part fit is employed, incorrect host matches will produce additional bias through incorrect host mass estimates, leading to different best fit values for SN luminosities. We observe that 34.3\% of our simulated SNe Ia with the wrong host match ``switch sides", i.e. the correct host is on one side of the mass step but the matched host is on the other. Given that the rate of mismatches averaged over 50 realizations is 1.7\% after selection cuts for our Ia-only sample, the overall fraction of SNe Ia that switch sides of the mass step is $\sim 0.6\%$. We assume that such a small percentage of the sample switching sides makes little impact and we do not pursue this aspect of the analysis further.

\subsection{Cosmological Biases with Photometric Classification}
\label{subsec:cosmo-Ia+CC}

Some photometric classifiers, such as SuperNNova (SNN), rely on SN redshift information to improve classification accuracy. To evaluate the impact of incorrect redshifts from host galaxy mismatches on the predictions from photometric classifiers, we jointly simulate SNe Ia, two types of peculiar SNe Ia, and two types of core collapse SNe: SNII, SNIbc, SNIax, and SNIa-91bg. Details on these simulations can be found in Section~\ref{subsec:sn-models} and Table~\ref{tbl:sim-cuts-Ia-cc}.

We tested 4 different photometric classifiers on our SNIa+CC simulations: the baseline perfect classification, SNN with redshift information (\snnz), SNN without redshift information (\snnnoz), and SCONE. SNN is typically used with SN redshift information and has been shown to produce highly accurate Ia vs. non-Ia classification results in this paradigm, so testing both redshift-dependent and redshift-independent configurations will show which effect is more detrimental to performance: incorrect redshift information or lack of redshift information altogether. SCONE uses SN lightcurves alone without the need for redshift information, so its predictions are not affected by host matching. For this analysis, we choose to define an SN Ia classification as $P_{\text{Ia}} \geq 0.5$.

Following the same approach as the Ia-only analysis, $w$ shifts were calculated for each photometric classifier by comparing two sets of identical simulations with and without mismatches. The results are shown in Table~\ref{tbl:Ia-cc-results}.


\begin{figure}
    \includegraphics[scale=0.38,trim={0cm 0cm 0cm 0cm}]{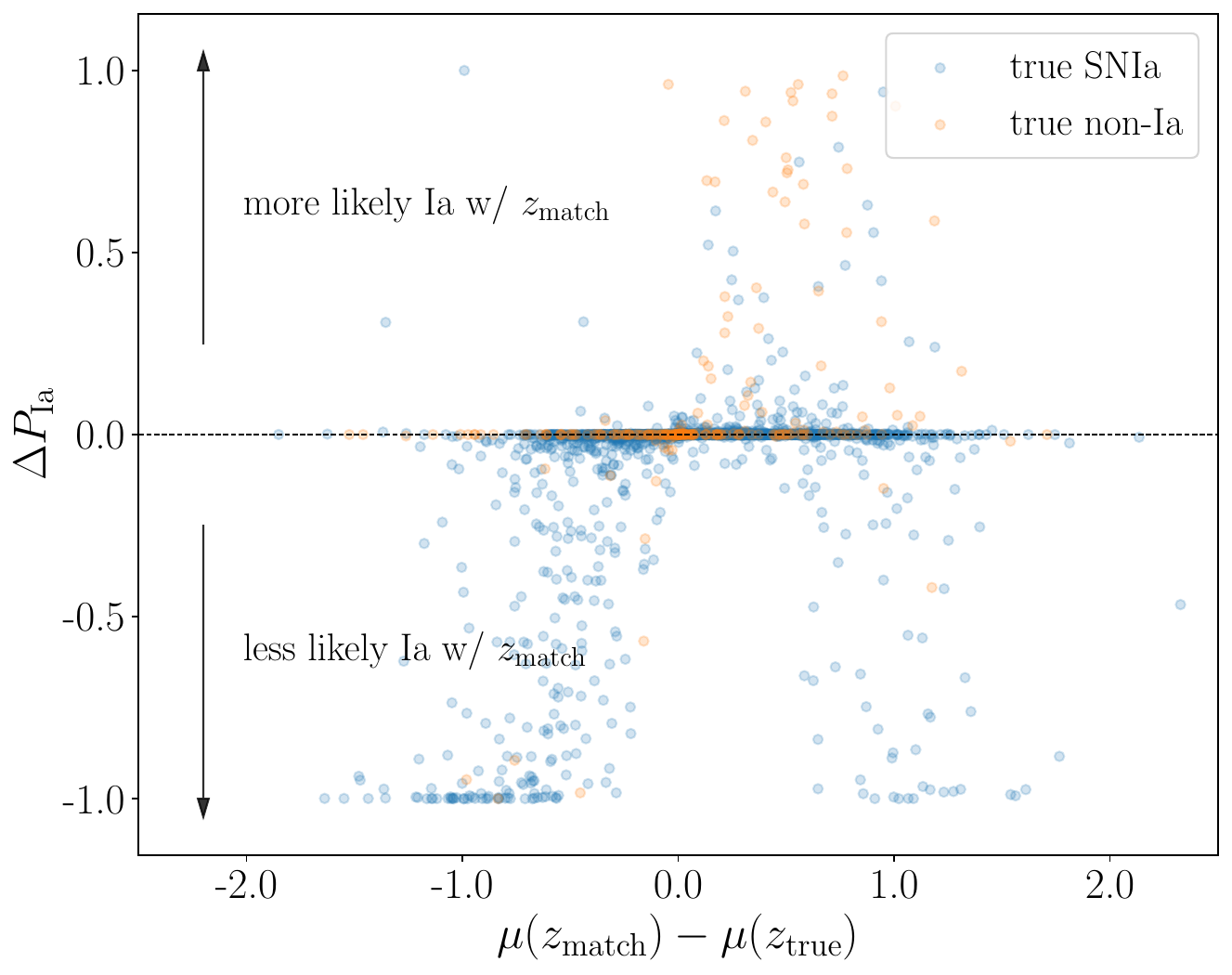}
    \centering
    \caption{Visualization of the impact of incorrect redshifts on \snnz\ predictions. $\Delta P_{\mathrm{Ia}}=P_{\mathrm{Ia, wronghost}} - P_{\mathrm{Ia, correcthost}}$ is the difference between $P_{\mathrm{Ia}}$ values output by the \snnz\ classifier given the wrong host redshift ($P_{\mathrm{Ia, wronghost}}$) and the correct host redshift ($P_{\mathrm{Ia, correcthost}}$) for SNe with mismatched hosts. $\Delta P_{\mathrm{Ia}}$ values are plotted against the difference between the distance modulus $\mu$ calculated at the wrong ($\mu(z_{\text{match}})$) and correct host redshifts ($\mu(z_{\text{true}})$). As $\mu(z_{\text{match}})-\mu(z_{\text{true}})$ deviate from 0, we would expect larger deviations in $P_{\mathrm{Ia}}$ values, i.e. $|\Delta P_{\mathrm{Ia}}| > 0$. Non-Ia SNe (orange points) more likely to be misclassified as Ia with the wrong redshift will appear in the upper half of the plot ($\Delta P_{\text{Ia}} > 0$), whereas SNe Ia (blue points) more likely to be misclassified as non-Ia with the wrong redshift will appear in the lower half. 7\% of mismatched SNe are incorrectly classified as a result of wrong host redshifts, leading to a overall 0.1\% reduction in classification accuracy compared to a simulation with correct host redshifts.}
    \label{fig:SNN_diffs}

\end{figure}

\begin{figure}
    \includegraphics[scale=0.38,trim={0cm 0cm 0cm 0cm}]{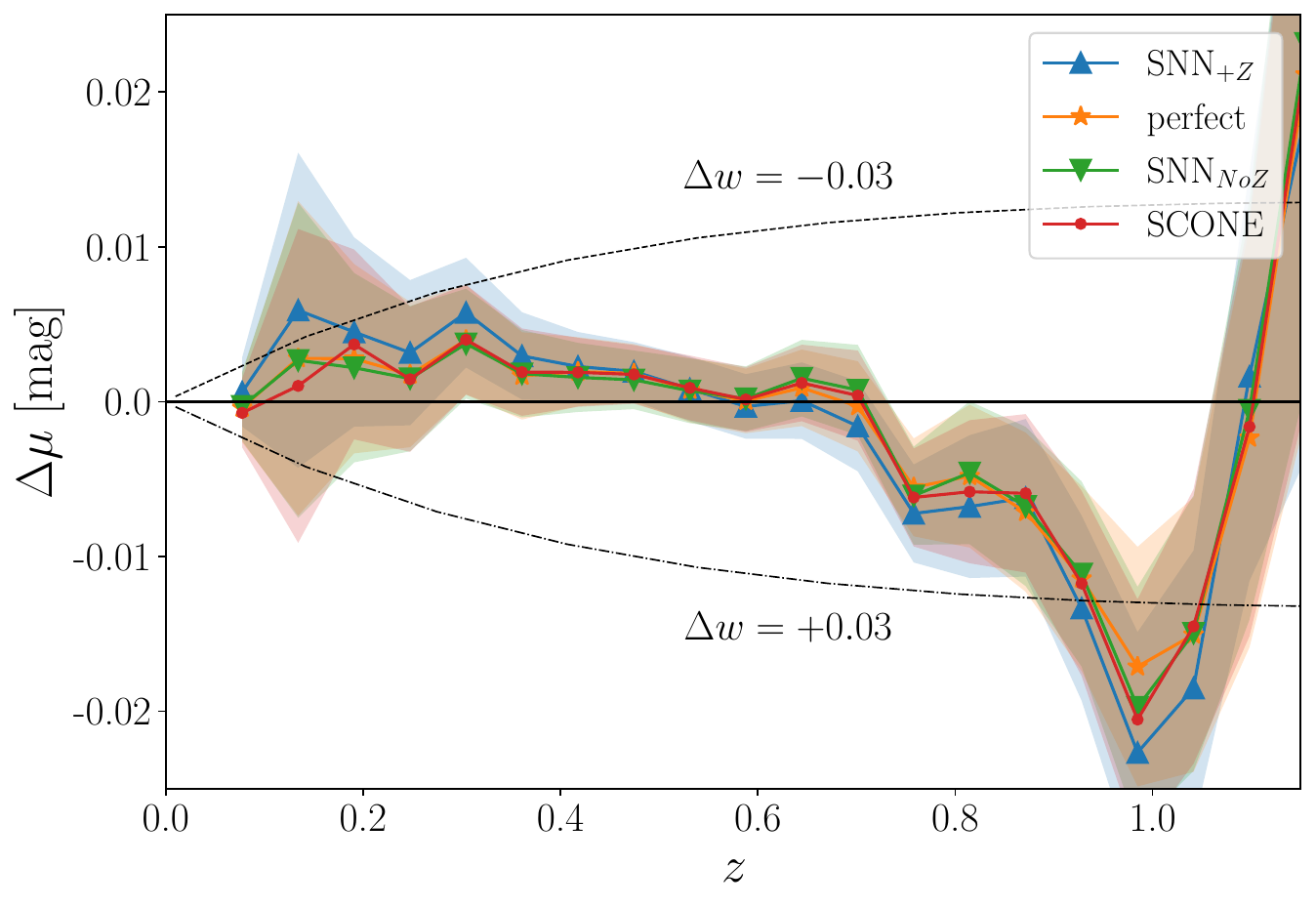}
    \centering
    \caption{Biases in binned Hubble residuals for all 4 photometric classifiers between samples with and without host galaxy mismatch. Uncertainties are shown as the shaded region and calculated from the binned standard deviations of $\mu_{\mathrm{match}}$ and $\mu_{\mathrm{truehost}}$. Lines showing $\Delta w = \pm 0.03$ are also plotted for reference.}
    \label{fig:Ia+CC_delta-mu}
\end{figure}

\begin{figure*}
    \includegraphics[scale=0.28,trim={0cm 0cm 0cm 0cm}]{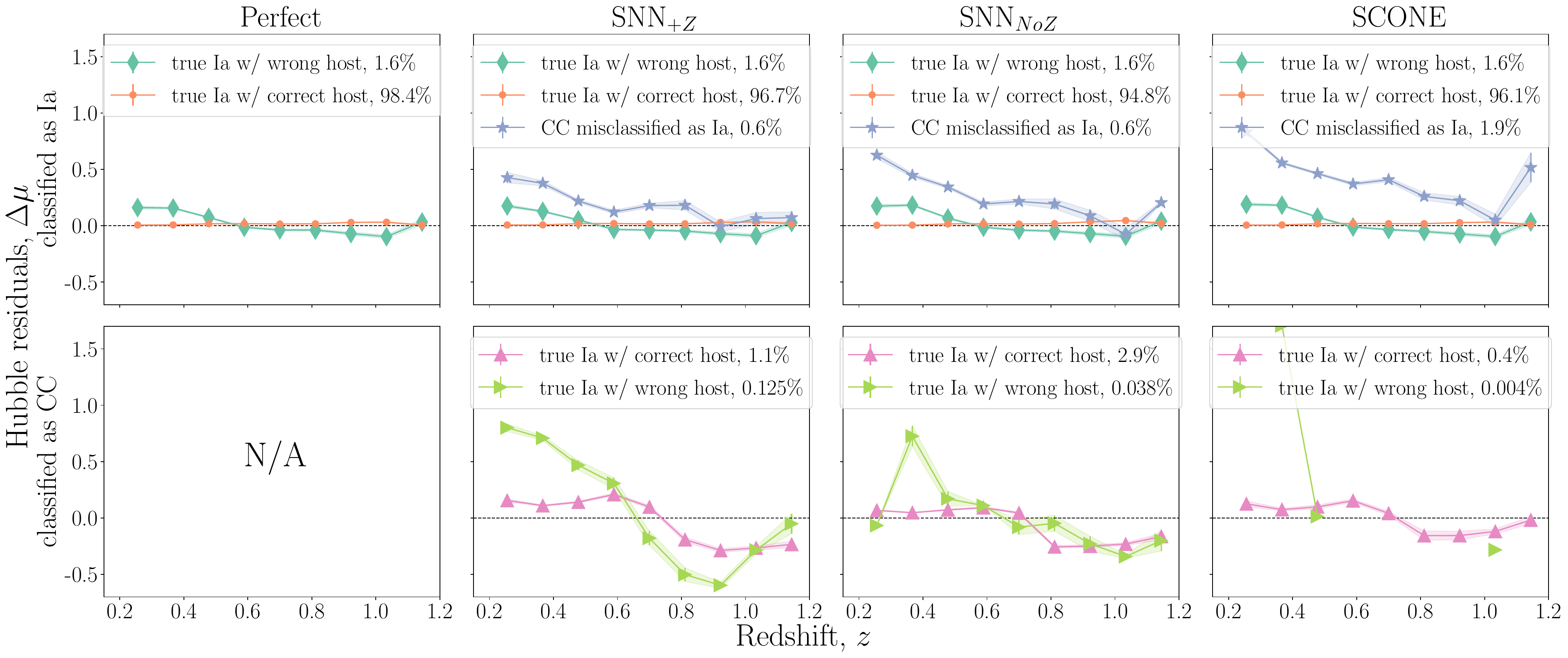}
    \centering
    \caption{Binned Hubble residuals, $\Delta \mu = \mu_{\mathrm{match}} - \mu_{\mathrm{model}}$, for the 4 photometric classifiers for our SNIa+CC simulation with mismatch. (top) SN populations classified as SNe Ia by each photometric classifier, including true SNe Ia with mismatched hosts as well as core collapse contamination. (bottom) SN populations classified as CC SNe, including SNe with mismatched hosts.}
    \label{fig:Ia+CC_residuals}
\end{figure*}
\snnz\ is the only classifier that should be impacted by host galaxy mismatches, and this relationship is observed in the larger $\Delta w$ value and lower classification accuracy for simulations with mismatched host galaxies. Figure~\ref{fig:SNN_diffs} shows the difference between predicted SN Ia probability output by the \snnz\ classifier given the wrong host redshift ($P_{\mathrm{Ia,wronghost}}$) as opposed to the correct host redshift  ($P_{\mathrm{Ia,correcthost}}$)}. Wrong redshifts indeed cause SNN to produce incorrect predictions for both Ia and non-Ia SNe, leading to the observed drop in accuracy with wrong redshifts. SCONE and \snnnoz\ are oblivious to any host matching changes and produce the same predictions for both sets of simulations, as expected. 

Figure~\ref{fig:Ia+CC_delta-mu} shows the biases on the binned Hubble diagram for each of the photometric classifiers with $\Delta \mu$ defined as in Equation~\ref{eq:delta-mu}. Overall, the classifiers perform quite similarly and exhibit very small differences in $\Delta \mu$ over the full redshift range. As expected from the small $\Delta w$ values, the $\Delta \mu$ curves for the Ia+CC simulations exhibit a slight redshift-dependent bias, though still mostly consistent with 0 up to high redshifts. Further validating the observed $\Delta w$ values for each classifier, we see that the two classifiers with most similar $w$ shifts, the perfect classifier (shown in orange) and \snnnoz\ (green), have the most similar $\Delta \mu$ values. \snnz\ (shown in blue), which has the largest $w$ shift, also consistently appears furthest from $\Delta \mu=0$ across all redshift bins.

The binned Hubble residuals of SNe Ia in the Ia+CC simulations as predicted by the 4 photometric classifiers are shown in Figure~\ref{fig:Ia+CC_residuals}. This plot shows $\Delta \mu = \mu_{\mathrm{match}} - \mu_{\mathrm{model}}$, as opposed to Figure~\ref{fig:Ia+CC_delta-mu}, which shows $\mu_{\mathrm{match}} - \mu_{\mathrm{truehost}}$. This allows us to compare residuals from the subpopulations of SNe with wrong and correct host match with respect to a fiducial cosmology. 
The biases on Hubble residuals from SNe Ia with wrong hosts (shown in teal diamonds on the top row) appear similar between the four classifiers, reflecting the small recovered $\Delta w$ values shown in Table~\ref{tbl:Ia-cc-results}. We also observe that the bias from wrong hosts is much more pronounced in SNe Ia misclassified as CC (shown in green triangles on the bottom row), particularly in the \snnz\ panel, indicating that \snnz\ was able to identify severe redshift outliers and rejected them from the SN Ia sample.

\subsection{Robustness of Cosmological Biases}
\subsubsection{Impact of CMB Prior}
The $w - \Omega_m$ contour estimated from measurements of the CMB exhibits a nearly orthogonal direction of degeneracy to the SN-only contour for a flat $w\text{CDM}$ model, providing strong constraints and drastically reducing the impact of systematics that act along the SN degeneracy direction. All $\Delta w$ values reported in Sections~\ref{subsec:cosmo-Ia} and~\ref{subsec:cosmo-Ia+CC} were calculated with a CMB prior. In this section, we evaluate the impact of the CMB prior on cosmological biases.

\begin{figure}
     \centering
     \includegraphics[scale=0.6,trim={1cm 0cm 0cm 0cm}]{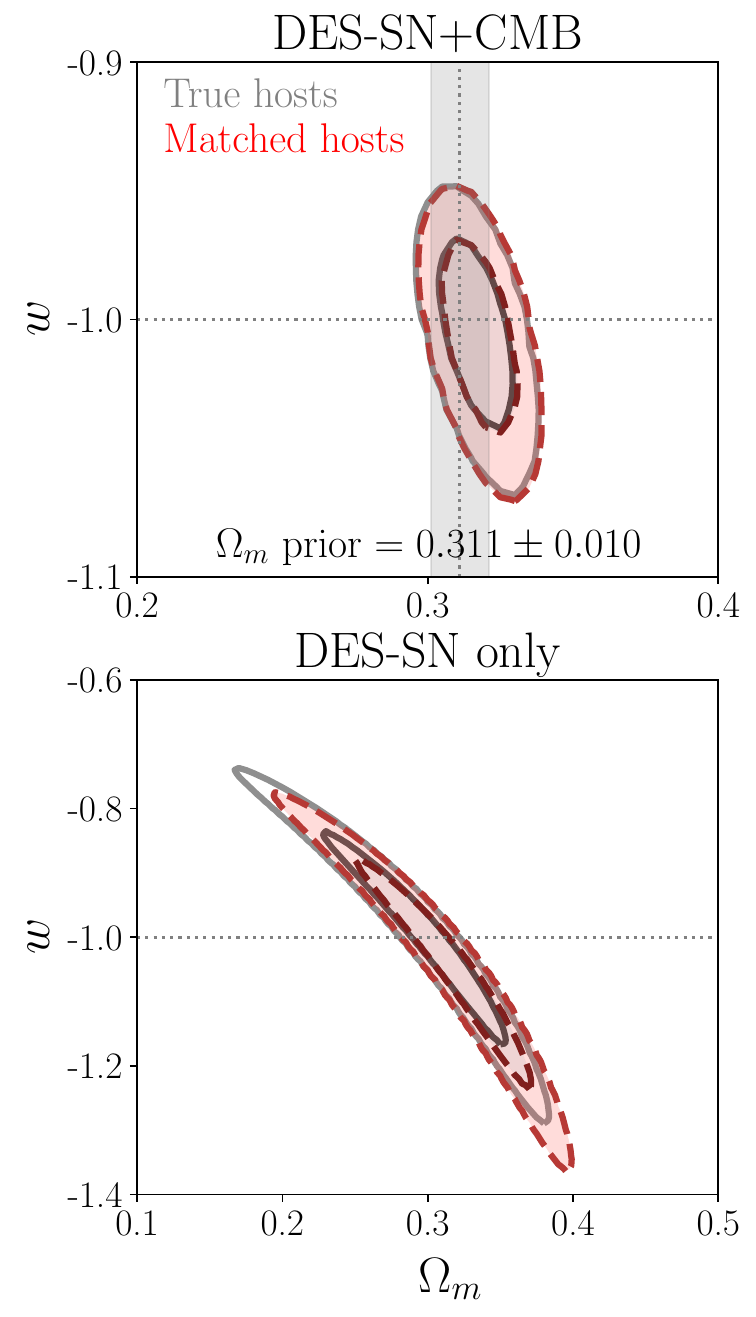}
    \caption{Cosmological contours computed with one realization of simulated DES Y5 SNe Ia with (red) and without host galaxy mismatch (gray). (top) SN+CMB contours, in which an approximate CMB prior $\Omega_m=0.311\pm 0.010$ prior is applied (see Section~\ref{subsec:cosmo-params} for details). This results in good agreement between the two contours. (bottom) SN-only contours showing a substantial shift in best fit $w, \Omega_m$ between the two contours. }
     \label{fig:contour}
     
 \end{figure}

The cosmological biases on both $w$ and $\Omega_m$ with and without the CMB prior are shown in Table~\ref{tbl:shifts-cmb} for the SNIa-only sample. The associated cosmological contours are shown in Figure~\ref{fig:contour}. Both $\Delta w$ and $\Delta \Omega_m$ are significantly inflated without the CMB prior, though still within their uncertainties (right column). The larger shift in $ w$ and $\Omega_m$ are visible when comparing the contours in Figure~\ref{fig:contour}. The contours computed with a CMB prior (top panel) are very nearly identical, whereas with a flat $\Omega_m$ prior, the contour with matched hosts is visibly shifted from the true hosts contour. However, these results are still consistent with 0 for the analysis we performed for the DES data, but should be studied further in future surveys.

\begin{table}
    \centering
    
    \begin{tabular}{c c c }
        \toprule
       Parameter & with CMB prior & no CMB prior\\
        \midrule
        $\Delta w$ & $0.0013 \pm 0.0026$ & $-0.062 \pm 0.072$ \\ 
        $\Delta \Omega_m$ & $0.0014 \pm 0.0017$ & $0.028 \pm 0.029$ \\
        \bottomrule
    \end{tabular}
    \caption{$\Delta w$ and $\Delta \Omega_m$ values for the Ia-only SN population with and without a CMB prior. The values in the $\Delta w$ with CMB prior cell are reproduced from Table~\ref{tbl:Ia-cc-results}.}
    \label{tbl:shifts-cmb}
\end{table}

    
\subsubsection{Bias Correction Simulations with Mismatch}
\label{subsec:bcor-mismatch}
In the earlier sections, distance moduli from both $S_{\mathrm{match}}$ and $S_{\mathrm{truehost}}$ are bias corrected using a large simulation of SNe Ia generated with the same parameters as $S_{\mathrm{truehost}}$, i.e. each SN is matched to its true host. We define this set of perfectly matched bias correction simulations as $B_{\mathrm{truehost}}$. An alternative bias correction strategy is to ``correct like with like", i.e. generating two separate bias correction simulations, one identical to $S_{\mathrm{match}}$ as well as the existing one identical to $S_{\mathrm{truehost}}$. We define this new set of DLR-matched bias correction simulations as $B_{\mathrm{match}}$. $B_{\mathrm{match}}$ models the biases arising from mismatched hosts in the bias correction simulations in order to correct for these biases in the simulated data. The reference $w$ value, $w_{\mathrm{truehost}}$ is still computed from $S_{\mathrm{truehost}}$ corrected with the baseline bias correction simulations with correct hosts only ($B_{\mathrm{truehost}}$). In this scheme, we define $w(S,B)$ as the best-fit $w$ value computed from simulations $S$ corrected with bias correction simulations $B$. The resulting $\Delta w$ equation then becomes
\begin{multline}
    \Delta w, \mathrm{bcor+mismatch}= \\\langle w(S_{\mathrm{match}}, B_{\mathrm{match}}) - w(S_{\mathrm{truehost}}, B_{\mathrm{truehost}}) \rangle_{\mathrm{(25\;realizations)}}.
\end{multline}

The $\Delta w$ values following this approach are shown in Table~\ref{tbl:w-shifts-bcor}. While the $\Delta w$ values using this bias correction scheme are still consistent with zero, the uncertainties are larger than those using bias correction simulations with perfect host matching; this may arise from comparing $w$ values corrected with two statistically independent sets of bias correction simulations, i.e. $S_\mathrm{match}$ is corrected with $B_\mathrm{match}$, whereas $S_{\mathrm{truehost}}$ is corrected with $B_\mathrm{truehost}$. Further investigation of the interplay of bias correction simulations with different sources of bias, including incorrect redshifts from host mismatch, will be addressed in a future work.

\begin{table}
    \centering
    \caption{$\Delta w$ values for the Ia-only and Ia+CC simulated SN populations with an alternative bias correction scheme (``bcor+mismatch") that includes mismatched hosts in the bias correction simulations. The values in the $\Delta w$, baseline column are reproduced from Table~\ref{tbl:Ia-cc-results}.}
    \label{tbl:w-shifts-bcor}
    \begin{tabular}{l c c c}
        \toprule
        Classifier & SNe  & $\Delta w$, baseline & $\Delta w$, bcor+mismatch \\
        \midrule
        Perfect & Ia only & $0.0013 \pm 0.0026$ & $-0.0094 \pm 0.0099$\\ 
        Perfect & Ia+CC & $0.0011 \pm 0.0027$ & $-0.0110 \pm 0.0120$ \\ 
        \snnz & Ia+CC & $0.0032 \pm 0.0040$ & $-0.0081 \pm 0.0086$ \\ 
        \snnnoz & Ia+CC & $0.0009 \pm 0.0028$ & $-0.0100 \pm 0.0012$ \\
        SCONE & Ia+CC & $0.0016 \pm 0.0032$ & $-0.0099 \pm 0.0100$ \\
        \bottomrule
    \end{tabular}
\end{table}


\subsubsection{Results from Varying Sérsic Scale}
\label{subsec:sersic-scale}
We vary the scaling of the fitted Sérsic $a$ and $b$ parameters, which describe each galaxy's semi-major and semi-minor axes, respectively. Simulated SNe are placed according to the intrinsic light profile described by these parameters. The scaled parameters $a'$ and $b'$ are calculated as $a'=ka, b'=kb$, where $k$ is the scaling parameter we vary. We tested $k \in [0.5, 1.2]$ with an interval of 0.1 and evaluated $\chi^2$ values on the histograms in Figure~\ref{fig:param-dists} to find the best match between data and simulations. We found that $k=0.8$ minimized the $\chi^2$ and was thus chosen for the main analysis, as described in Section~\ref{subsec:catalog-cuts-params}. We note that our conclusion runs contrary to that of \citet{Li_2016}, which found that the Sérsic effective radius, $R_e = \sqrt{ab}$, is underestimated for stacked galaxy images rather than overestimated, as we discovered.

We performed cosmological parameter estimation using simulations with and without Sérsic scaling ($k=0.8$ and $k=1$) and found a modest benefit of using the $k=0.8$ scaling for redshift-independent photometric classifiers as well as perfectly classified SNe Ia, but a noticeable improvement for SNe classified using the redshift-dependent classifier, \snnz. This seems to indicate that the Sérsic scaling improves the host matching efficiency significantly, since \snnz\ is most affected by incorrect redshifts. This is notably the same scaling factor found to best match the DES3YR data when comparing distributions of host galaxy surface brightness at the SN position \citep[see Figure 6 of][]{des3yr}. The recovered biases on $w$ from mismatched hosts ($\Delta w$) with and without Sérsic scaling for all simulations and classifiers are shown in Table~\ref{tbl:w-shifts-sersic}.

\begin{table}
    \centering
    \caption{$\Delta w$ values for the Ia-only and Ia+CC simulated SN populations with and without Sérsic scaling ($k=0.8$ and $k=1$). The primary results presented in this work (Sections~\ref{subsec:cosmo-Ia} and~\ref{subsec:cosmo-Ia+CC}) use simulations scaled with $k=0.8$. The \snnz\ results with $k=1$ have an inflated $\Delta w$ as well as $\sigma_{\Delta w}$ due to a realization with poor $\chi^2$ fit from \texttt{wfit}.}
    \label{tbl:w-shifts-sersic}
    \begin{tabular}{l c c c}
        \toprule
        Classifier & SNe & $\Delta w(k=0.8)$ & $\Delta w(k=1)$ \\
        \midrule
        Perfect & Ia only & $0.0013 \pm 0.0026$ & $0.0030 \pm 0.0042$\\ 
        Perfect & Ia+CC & $0.0011 \pm 0.0027$ & $0.0016 \pm 0.0026$ \\ 
        \snnz & Ia+CC & $0.0032 \pm 0.0040$ & $0.0120 \pm 0.0640$ \\ 
        \snnnoz & Ia+CC & $0.0009 \pm 0.0028$ & $0.0021 \pm 0.0031$ \\
        SCONE & Ia+CC & $0.0016 \pm 0.0032$ & $0.0025 \pm 0.0034$ \\
        \bottomrule
    \end{tabular}
\end{table}

\subsubsection{Host Confusion Parameter}
\begin{figure}
    \includegraphics[scale=0.36]{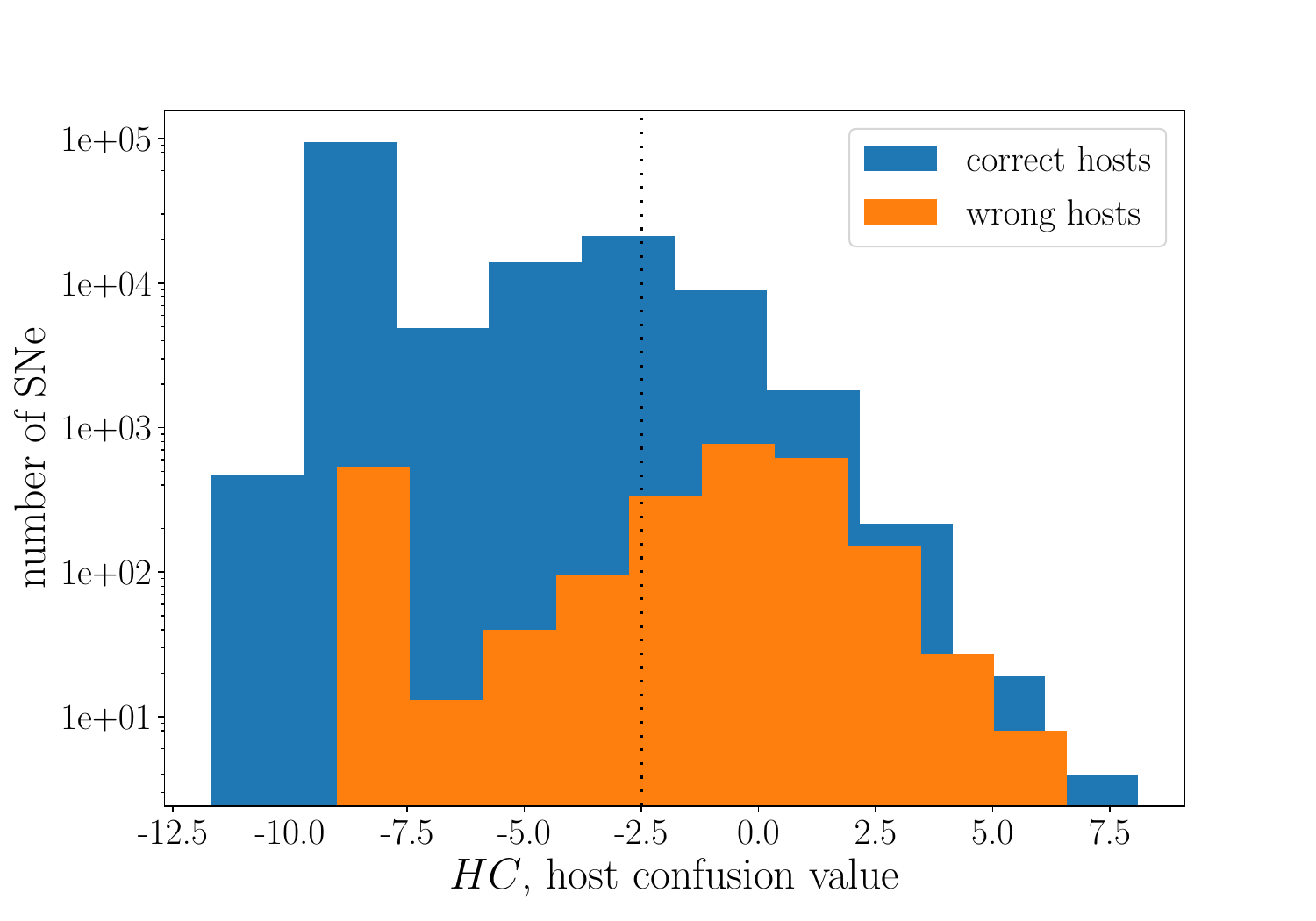}
    \centering
    \caption{Distributions of $HC$ values for SNe with wrong and correct host matches. The dotted line is drawn at our chosen threshold of $HC = -2.5$.}
    \label{fig:host-confusion-dists}
\end{figure}

Equation 3 in \cite{gupta} defines a quantity characterizing the likelihood of a wrong match: the host confusion parameter, or $HC$. This parameter is a function of the \ddlr values of each galaxy in the search radius around a SN, and aims to distinguish situations with a clear correct host from those without. In this analysis, we attempt to remove wrong hosts by cutting out SNe with high host confusion ($HC \geq -2.5$). The distribution of $HC$ values for wrong and correct hosts in our simulations is shown in Figure~\ref{fig:host-confusion-dists}. The $-2.5$ threshold was chosen by visual inspection of this distribution. The $HC$ distributions for both populations look quite similar, but 72\% of SNe with an incorrect host match are removed by the $HC \geq -2.5$ cut, while 12\% of SNe with correct host matches are removed. 

We performed a full cosmology analysis with perfectly classified Ia-only simulations and find $\Delta w = -0.0043 \pm 0.0047$ by comparing Ia populations with and without wrong hosts, both subject to the $HC < -2.5$ selection requirement. In this cosmology analysis, we use bias correction simulations with host matching (i.e. $B_{\mathrm{match}}$ from Section~\ref{subsec:bcor-mismatch}) and select only the subset of the bias correction simulations with the same $HC < -2.5$ requirement for consistency. Both simulations (with perfectly matched hosts and with wrong hosts) are bias corrected with $B_{\mathrm{match}}$ in order to apply the $HC$ selection requirement globally, but it may not be suitable to correct simulations without wrong hosts in this way. The magnitude of this $\Delta w$ value is larger than that of the baseline (i.e. no $HC$ selection requirement), likely due to our inclusion of wrong hosts in the bias correction simulations, but is still consistent with zero.

\begin{figure}
    \includegraphics[scale=0.55]{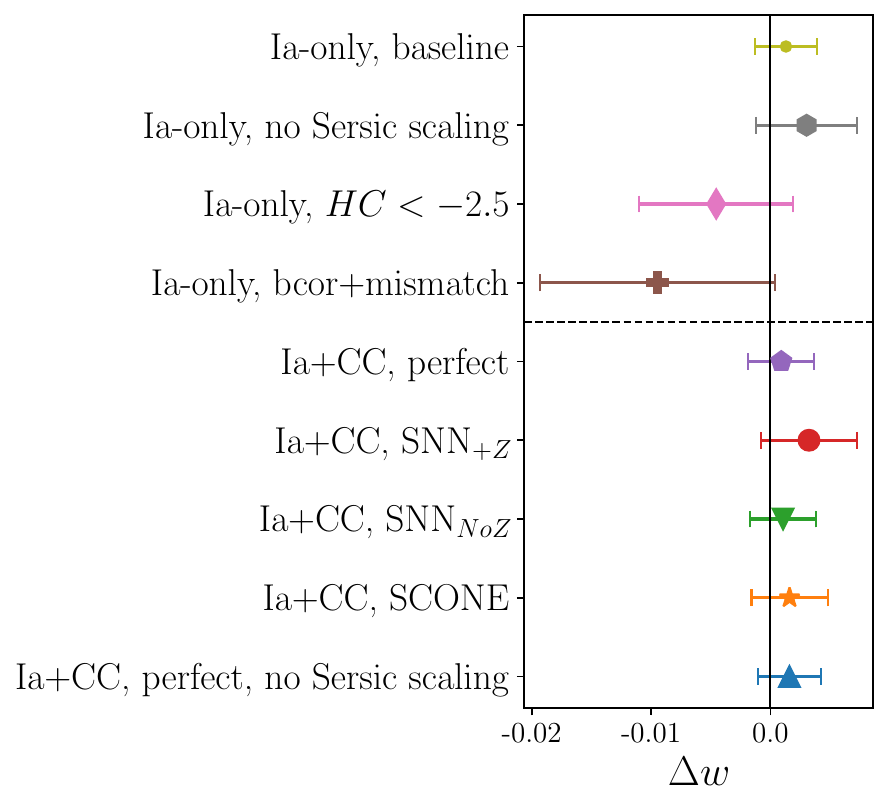}
    \centering
    \caption{$\Delta w$ values from all tested variants of simulated DES SNe with CMB prior. Note that the total uncertainty on $w$ from the DES3YR analysis \citep{des3yr_cosmo}, $\delta w = 0.059$, is much larger than the bounds of this plot. Details about each variant can be found in Section~\ref{sec:results}.}
    \label{fig:w_shifts}
\end{figure}

\section{Conclusions}

Matching SNe to their host galaxies is a non-trivial problem when using 2-dimensional images of 3-dimensional space. Accurate host galaxy matches are important for cosmology because host galaxies are used to measure the vast majority of SN redshifts. Identifying incorrect host galaxies alters the shape of the Hubble diagram and can impact the resulting fitted cosmological parameters.

In this work, we investigated the impact of mismatched host galaxies for the DES Y5 SN cosmology analysis. To this end, we created a galaxy catalog from DES deep field images that was sufficiently deep and dense, and calculated photometric redshifts, galaxy parameters, as well as Sérsic parameter fits for these 4 million galaxies. Simulations using this galaxy catalog were verified to be sufficiently similar to the DES Y5 photometric SN Ia sample using distributions such as Figure~\ref{fig:param-dists}. This simulation enabled us to predict the prevalence of mismatched host galaxies in DES data and, in turn, characterize the effect of host mismatches on cosmological parameter estimates.

Host matching was performed using the directional light radius (DLR) method, a  technique that identifies the host galaxy by minimizing SN-galaxy distance normalized by galaxy radius. This analysis aims to characterize the systematic error due to host misidentification by the DLR method for the DES Y5 cosmology analysis.

The main findings of this work are summarized in Figure~\ref{fig:w_shifts}, which shows the observed shifts in the dark energy equation of state parameter $w$ as a result of host galaxy mismatches with different approaches to the analysis. We defined $\Delta w$ by producing two identical sets of simulations: one with perfect host matching and one with host matches from the DLR method. We also probed the interplay between host galaxy mismatches and photometric classification, as classifiers such as SuperNNova use SN redshift estimates for more accurate SN type predictions. Finally, we explored the impact of variations to our analysis, such as the choice of $\Omega_m$ prior, bias correction simulations, S{\'e}rsic scaling factor, and additional selection cuts to remove incorrect host matches.

We found that the baseline $w$ shift with perfectly classified type Ia SNe is $\Delta w = 0.0013 \pm 0.0026$, and changes in certain properties can increase this to $| \Delta w | \sim 0.004$. We also find that the choice of photometric classifier makes an impact on the $w$ shift: classifiers requiring redshift estimates for prediction tend to misclassify SNe with the wrong redshift, leading to a larger $w$-bias. When the CMB prior is replaced with a flat $\Omega_m$ prior, the $\Delta w$ value changes to $-0.062 \pm 0.072$. Though the $\Delta w$ uncertainty is larger with the flat $\Omega_m$ prior, the shift is still consistent with 0 given the associated inflation in uncertainty. 

In conclusion, we find that our current estimate on the systematic error associated with host galaxy mismatch is substantially smaller than the statistical error for DES Y5, but as future surveys continue to discover more SNe Ia, we encourage continued study and improvement of the accuracy of host galaxy matching and prevalence of catastrophic errors in redshift.

\section*{Acknowledgements}
H.Q., J.L., and M.S. were supported by DOE grant DE-FOA-0002424 and NSF grant AST-2108094.
L.G. acknowledges financial support from the Spanish Ministerio de Ciencia e Innovaci\'on (MCIN), the Agencia Estatal de Investigaci\'on (AEI) 10.13039/501100011033, and the European Social Fund (ESF) ``Investing in your future" under the 2019 Ram\'on y Cajal program RYC2019-027683-I and the PID2020-115253GA-I00 HOSTFLOWS project, from Centro Superior de Investigaciones Cient\'ificas (CSIC) under the PIE project 20215AT016, and the program Unidad de Excelencia Mar\'ia de Maeztu CEX2020-001058-M.
PW acknowledges support from the Science and Technology Facilities Council (STFC) grant ST/R000506/1.
L.K. thanks the UKRI Future Leaders Fellowship for support through the grant MR/T01881X/1. 

This work was completed in part with resources provided by the University of Chicago’s Research Computing Center, as well as resources of the National Energy Research Scientific Computing
Center (NERSC), a DOE Office of Science User Facility
supported by the Office of Science of the U.S. Department of
Energy under Contract No. DE-AC02-05CH11231. 

Funding for the DES Projects has been provided by the U.S. Department of Energy, the U.S. National Science Foundation, the Ministry of Science and Education of Spain, 
the Science and Technology Facilities Council of the United Kingdom, the Higher Education Funding Council for England, the National Center for Supercomputing 
Applications at the University of Illinois at Urbana-Champaign, the Kavli Institute of Cosmological Physics at the University of Chicago, 
the Center for Cosmology and Astro-Particle Physics at the Ohio State University,
the Mitchell Institute for Fundamental Physics and Astronomy at Texas A\&M University, Financiadora de Estudos e Projetos, 
Funda{\c c}{\~a}o Carlos Chagas Filho de Amparo {\`a} Pesquisa do Estado do Rio de Janeiro, Conselho Nacional de Desenvolvimento Cient{\'i}fico e Tecnol{\'o}gico and 
the Minist{\'e}rio da Ci{\^e}ncia, Tecnologia e Inova{\c c}{\~a}o, the Deutsche Forschungsgemeinschaft and the Collaborating Institutions in the Dark Energy Survey. 

The Collaborating Institutions are Argonne National Laboratory, the University of California at Santa Cruz, the University of Cambridge, Centro de Investigaciones Energ{\'e}ticas, 
Medioambientales y Tecnol{\'o}gicas-Madrid, the University of Chicago, University College London, the DES-Brazil Consortium, the University of Edinburgh, 
the Eidgen{\"o}ssische Technische Hochschule (ETH) Z{\"u}rich, 
Fermi National Accelerator Laboratory, the University of Illinois at Urbana-Champaign, the Institut de Ci{\`e}ncies de l'Espai (IEEC/CSIC), 
the Institut de F{\'i}sica d'Altes Energies, Lawrence Berkeley National Laboratory, the Ludwig-Maximilians Universit{\"a}t M{\"u}nchen and the associated Excellence Cluster Universe, 
the University of Michigan, NSF's NOIRLab, the University of Nottingham, The Ohio State University, the University of Pennsylvania, the University of Portsmouth, 
SLAC National Accelerator Laboratory, Stanford University, the University of Sussex, Texas A\&M University, and the OzDES Membership Consortium.

Based in part on observations at Cerro Tololo Inter-American Observatory at NSF's NOIRLab (NOIRLab Prop. ID 2012B-0001; PI: J. Frieman), which is managed by the Association of Universities for Research in Astronomy (AURA) under a cooperative agreement with the National Science Foundation.

The DES data management system is supported by the National Science Foundation under Grant Numbers AST-1138766 and AST-1536171.
The DES participants from Spanish institutions are partially supported by MICINN under grants ESP2017-89838, PGC2018-094773, PGC2018-102021, SEV-2016-0588, SEV-2016-0597, and MDM-2015-0509, some of which include ERDF funds from the European Union. IFAE is partially funded by the CERCA program of the Generalitat de Catalunya.
Research leading to these results has received funding from the European Research
Council under the European Union's Seventh Framework Program (FP7/2007-2013) including ERC grant agreements 240672, 291329, and 306478.
We  acknowledge support from the Brazilian Instituto Nacional de Ci\^encia
e Tecnologia (INCT) do e-Universo (CNPq grant 465376/2014-2).

This manuscript has been authored by Fermi Research Alliance, LLC under Contract No. DE-AC02-07CH11359 with the U.S. Department of Energy, Office of Science, Office of High Energy Physics.

\bibliography{references}
\end{document}